\documentclass[a4paper,fleqn,usenatbib]{mnras}
\usepackage{newtxtext,newtxmath}
\usepackage[T1]{fontenc}
\usepackage{ae,aecompl}

\usepackage{bm} 
\usepackage{amsmath} 
\usepackage{amssymb} 
\usepackage{cases} 
\usepackage{graphicx}

\usepackage[dvipsnames]{xcolor}
\usepackage[normalem]{ulem}

\title[The Perturbed FLRW metric on all scales]{The perturbed FLRW metric on all scales: Newtonian limit and top-hat collapse}

\author[A. Finke]{Andreas Finke,$^{1}$\thanks{E-mail: andreas.finke@unige.ch}
\\
$^{1}$D\'epartement de Physique Th\'eorique and Center for Astroparticle Physics, Universit\'e de Gen\'eve, 24 quai Ansermet, 1211 Gen\'eve 4, CH}

\pubyear{2019}

\newcommand{\changed}[1]{{#1}}
\newcommand{\changedagain}[1]{{{#1}}}
\newcommand{\changedagainagain}[1]{{{#1}}}

\begin{document}
\label{firstpage}
\pagerange{\pageref{firstpage}--\pageref{lastpage}}
\maketitle

\begin{abstract}

The applicability of a linearized perturbed FLRW metric to the late, lumpy universe has been subject to debate. \changed{
	We consider in an elementary way the Newtonian limit of the Einstein equations with this ansatz for the case of structure formation in late-time cosmology, on small and on large scales, and argue that linearizing the Einstein tensor produces only a small error down to arbitrarily small, decoupled scales (e.g.~solar system scales). 
	On subhorizon patches, the metric scale factor becomes a coordinate choice equivalent to choosing the spatial curvature, and \emph{not} a sign that the FLRW metric cannot perturbatively accommodate very different local physical expansion rates of matter; we distinguish these concepts, and show that they merge on large scales for the Newtonian limit to be globally valid. Furthermore, on subhorizon scales, a perturbed FLRW metric ansatz does not already imply assumptions on isotropy, and effects beyond a FLRW background, including those potentially caused by nonlinearities of GR, may be encoded into nontrivial boundary conditions. The corresponding cosmologies have already been developed in a Newtonian setting by Heckmann and Sch\"ucking and none of these boundary conditions can explain the accelerated expansion of the universe. Our analysis of the field equations is confirmed on the level of solutions by an example of pedagogical value, comparing a collapsing top-hat overdensity (embedded into a cosmological background) treated in such perturbative manner to the corresponding exact solution of GR, where we find good agreement even in the regimes of strong density contrast.
}

\end{abstract}


\section{Introduction}

The late universe, a slice of which is visible to galaxy surveys on our past lightcone, is very clumpy. This is in stark contrast to the earlier universe: the observed fluctuations in the cosmic microwave background CMB are small~\citep{planck2018}, allowing for a precise interpretation of their properties based on linearized, relativistic cosmological perturbation theory that is well established~\citep{durrer2008CMB}. However, gravitational collapse has transformed the smooth state into one that cannot be evolved anymore by effective fluid equations~\citep{baumann2012EFTLSS} without removing short scales of comoving size of a couple of Mpc, and certainly not by linear equations. 

The conceptually simple, but technically challenging, nonlinear Newtonian Vlasov-Poisson system~\citep{peebles1980large} describes the gravitational amplification of dark matter structures including smaller scales and large density contrasts. Newtonian N-body simulations have tackled the challenging prediction of the structures' morphology, solving this system by sampling its characteristic curves. \changed{There is no doubt that Newtonian theory is capturing the gravitational physics of finite astrophysical systems such as galaxies or clusters of galaxies, where gravity is weak. We similarly expect Newtonian physics to approximate sufficiently localized cosmological patches when fields are weak  and matter is moving slowly~\citep{peebles1980large, kaiser2017NewtonianBackreaction}, including the nonlinear evolution of matter beyond linear cosmological perturbation theory. However, in cosmology, we are interested in global solutions, and the validity and generality of a Newtonian approximation is not immediately clear when going to large scales. }

The Newtonian limit of general relativity (GR) may therefore be studied carefully in the context of large-scale structure in cosmology~\citep{peebles1980large, holz1998Lensing, greenwald2012newtonian}. 
\changed{The limit involves linearizing the Einstein tensor for weak fields around a background. The choice of this global background solution is nontrivial. Instead, a local approach aims to avoid this step and it has been shown how gluing together patches close to the Minkowski spacetime in a bottom-up construction yields cosmological solutions of GR~\citep{sanghai2015}, that is, the background emerges and can be compared to the predictions of Newtonian cosmology. 
Therefore, a local approach can help clarifying the generality of the Newtonian limit in cosmology, which is crucial when building simple models for reaching robust conclusions about our universe comparing to observations. For example, the report in~\cite{racz2017nbody} of large relativistic acceleration effects beyond Newtonian cosmology without the need for dark energy produced in simulations with a modified N-body code has recently received a critical reply~\cite{kaiser2017NewtonianBackreaction} in which the backreaction proposal (see e.g.~\cite{buchert2012backreaction}) was discussed assuming a Newtonian setting. However, any significant differences of Newtonian and GR cosmological dynamics claimed to invalidate this reply~\citep{buchert2017backreaction} must be related to assumptions made when taking the Newtonian limit of GR, and the discussion can be supplemented by this step.
Beyond foundational questions in GR, a Newtonian limit is still the foundation for obtaining the leading-order corrections to structure formation from covariantly formulated modified gravity theories, by using N-body simulations that have been modified accordingly. }
Furthermore, the Newtonian limit is the relevant approach for properly addressing cosmological effects on cosmologically very small systems, like the solar system (or even an atom~\citep{bonnor1999atom, price2012expanding}), both for GR~\citep{dicke1964expansion,bonnor1996expansion, cooperstock1998influence, carrera2010influence, nandra2012lambda} and modified gravity. Constraints on the time evolution of the Newton constant obtained in the solar system~\citep{pitjeva2013, hofmann2018LLR, fienga2015} and therefore on modified gravity are already below one percent of the Hubble rate, and are becoming stronger with observation time with dramatic implications for some models~\citep{me2019LLR}. 

In this paper, we are concerned with a correct leading-order treatment of nonlinear structures in late-time cosmology. This is complemented by recent works on more general expansions of GR for a highly accurate description of our universe~\citep{adamek2013perturb, adamek2016gevolution, fidler2017weakfieldNbody, milillo2015link,goldberg2017allscales}.

We base the discussion on a convenient standard tool, which is the FLRW metric with scalar and vector perturbations~\citep{harrison1967normal, bertschinger1995notes} 
\begin{align}
\label{eq:lineelement}
ds^2 = -(1+2\Psi) \mathrm{d}t^2 + 2 \mathbf{B} \cdot\mathrm{d} \boldsymbol{x}  \mathrm{d} t  + a^2(t)(1-2\Phi)\mathrm{d}\boldsymbol{x}^2,
\end{align}
where $\Psi$, $\Phi$ and $\mathbf{B}$ are small in a way to be specified. 
The presence of the scale factor $a(t)$ at an early stage of the theoretical description is intriguing.
In contrast, in Newtonian cosmology, the starting point (before the continuum limit and without a cosmological constant) are simply the equations 
\begin{align}
\label{eq:Newton}
\ddot{\boldsymbol{r}}_i = \; G_N \, m\sum \frac{\boldsymbol{r}_j-\boldsymbol{r}_i}{|\boldsymbol{r}_j-\boldsymbol{r}_i|^3} 
\end{align}
for particles of equal mass $m$. The scale factor can now appear in two familiar, but in principle different, ways. The first situation concerns the dynamics of a homogeneous ball of dust in the continuum limit, of finite or infinite size, for which the time dependence of the radial positions of the shells is a common factor. Here, the scale factor is related to the physical motion of matter, and it satisfies a Friedmann equation. The second situation is that of an inhomogeneous cosmos. Here the scale factor is introduced into the Newtonian equations~(\ref{eq:Newton}) by hand, adopting comoving spatial coordinates $\boldsymbol{x}$ with  
\begin{align}
\boldsymbol{r}=a(t)\boldsymbol{x},
\end{align} 
\changed{which is a coordinate gauge that has been discussed carefully in}~\citep{kaiser2017NewtonianBackreaction}. \changed{As noted there, the most convenient, but by no means necessary, choice is to use the motion scale factor of the first situation that emerges when replacing the inhomogeneous cosmos by an averaged homogeneous ball. }

\changed{On the other hand, it is usually stated that the metric scale factor in~(\ref{eq:lineelement}) should be the one of a best-fitting background universe. When perturbations in~(\ref{eq:lineelement}) are neglected ``on average'', and if the background energy density is assumed to be the Newtonian spatial average of the physical density, the metric scale factor obeys the same Friedmann equation of the homogeneous ball of matter in Newtonian theory. 
Since the Einstein equations are nonlinear, a background scale factor based on Newtonian averaging may seem as a shortcut lacking rigorous justification}~\citep{kolb2010backreaction, wiltshire2011dust}. 
Furthermore, if the ``correct'' scale factor must be chosen in~(\ref{eq:lineelement}), corresponding to some average Hubble flow, the intuition may be that situations where matter locally moves with very different expansion rates (e.g.~after turnaround) cannot be small perturbations as assumed in~(\ref{eq:lineelement})~\citep{rasanen2010FLRW}.
However, it has also been shown that Lema\^{i}tre--Tolman--Bondi (LTB) solutions with locally different expansion rates of matter can be mapped to perturbed FLRW if the inhomogeneities are confined to subhorizon scale~\citep{karel2008LTBFLRW, yamamoto2016LTBFLRW}. This supports the validity of the ansatz~(\ref{eq:lineelement}) but clearly questions the physical relevance of the scale factor in the metric on small scales, \changed{similarly to its introduction as a coordinate choice in Newtonian cosmology}.

\changed{We therefore studied GR cosmology using a perturbed FLRW metric and guided by the observed properties of dark matter at low redshifts, emphasizing a local approach.  To leading order, we recovered the validity of a linearization of the Einstein tensor and obtained a consistent Newtonian limit.
 These results can be illustrated by an explicit example: the simplest, most extreme (and therefore most interesting) spacetime describing a collapsing object within the expanding universe is arguably that of a top-hat profile -- a homogeneous core surrounded by an empty shell -- embedded into a matter-dominated universe. We confirmed that the perturbed FLRW metric is able to describe this situation; it can thus simultaneously accommodate structures with very different expansion rates. As a key step in the argument, we derived explicit coordinate transformations that can change the scale factor nonperturbatively from being solution of one Friedmann equation to a solution of another Friedmann equation, while keeping the potentials perturbative. 
}
\changed{
We also found that on subhorizon scales and allowing for $\Psi \neq \Phi$ the scale factor in the metric~(\ref{eq:lineelement}) is arbitrary, whereas it emerges on large scales for a correct description of the geometry, where it eventually must correspond to the physical motion of matter. We checked this gauge freedom of $a(t)$ by repeating the analysis of the top-hat example using the perturbed FLRW metric with another choice of scale factor.  }

\changed{
 Due to the local and constructive approach, the calculations are general enough to have implications for the backreaction proposal. On small scales, where the scale factor of the perturbed FLRW metric is arbitrary, the effect of GR solutions is constrained free of the assumption that the global background is a FLRW spacetime. Since local cosmological tests take place in this regime, dark energy can be probed disentangled from potential relativistic effects on the dynamics. 
}

\changed{
The paper is structured as follows. Part~\ref{sec:newtonian} reviews the Newtonian limit for late-time cosmology and derives the starting point of Newtonian cosmology from GR in the general setting of arbitrary scale factor.  
Part~\ref{sec:tophat} then introduces the top-hat spacetime in detail in Section~\ref{sec:tophateqns}, tests the FLRW metric in the standard case in Section~\ref{sec:cosmoscalefactor}, and subsequently confirms the result of part~\ref{sec:newtonian}, for $\Psi \neq \Phi$ with a nonstandard choice of scale factor in Section~\ref{sec:collapsescalefactor}. Before we conclude in Section~\ref{sec:conclusion} we discuss in Section~\ref{sec:discussion} in some detail the large-scale limit (Section~\ref{sec:emergenceofa}), the meaning of cosmic expansion for local systems (Section~\ref{sec:expansionforce}), the transition from covariance to inertial frames (Section~\ref{sec:inertialframes}), and the assumptions that have been made with the FLRW metric and the implications for the backreaction proposal (Section~\ref{sec:backreaction}). }

\section{Newtonian limit}
\label{sec:newtonian}

We start from GR with cosmological constant $\Lambda$ and the metric~(\ref{eq:lineelement})
\begin{align}
ds^2 = -(1+2\Psi) \mathrm{d}t^2 + 2 \mathbf{B} \cdot\mathrm{d} \boldsymbol{x}  \mathrm{d} t  + a^2(t)(1-2\Phi)\mathrm{d}\boldsymbol{x}^2 
\end{align}
without spherical symmetry, but neglecting tensor perturbations. \changedagain{One may think about this ansatz as being perturbative around a flat FLRW background, but this is an overly restrictive picture on small enough scales. We will shortly see that a ``curvature'' term in the equation of motion for $a(t)$ can be encoded into a quadratic difference of $\Psi$ and $\Phi$, corresponding to a Friedmann equation for universes with nonzero spatial curvature. This is a gauge freedom appearing when integrating the Einstein equations carefully. It implies that there must exist corresponding coordinate transformations that leave the form of the metric~(\ref{eq:lineelement}) invariant but that change the scale factor nonperturbatively by $\mathcal{O}(1)$. These transformations preserve the right perturbative character of the potentials for late-time structure in cosmology. We derive them in for the top-hat example in Part~\ref{sec:tophat} (equations (\ref{eq:trafo1}, \ref{eq:trafo2}, \ref{eq:trafo3}, \ref{eq:trafo4}) and  (\ref{eq:trafo12}, \ref{eq:trafo22}, \ref{eq:trafo32}, \ref{eq:trafo42})). 
With a gauge change of $a(t)$ for a fixed spacetime the interpretation of the ``comoving'' spatial coordinates $ \boldsymbol{x}$ changes as well. For this reason these spatial coordinates  are not necessarily aligned with the motion of matter. Locally, such a construction would indeed be ill-defined in late-time cosmology because matter trajectories can cross; but even on average, some gauge freedom exists (see below around equation~(\ref{eq:subhorizonregime})).} 

The source on the right hand side of the Einstein equations, 
\begin{align}
G_{\mu\nu} + \Lambda g_{\mu\nu} = 8 \pi G_N T_{\mu\nu},
\end{align} can conveniently be thought of as a collection of point masses that sample an initial phase-space sheet which is allowed to cross itself in real space. For simplicity, we neglect pressure. Then, matter motion is geodesic. We avoid both a fluid assumption and averaging. For completeness, in the continuum limit an explicit expression for the energy-momentum tensor is given by
\begin{align}
T^{\mu \nu}(x^\alpha) = \int \left[dp^1 dp^2 dp^3 \frac{\sqrt{-g}}{ p_0 } \right] f (x^\alpha, p^i)  p^\mu p^\nu
\end{align}
where the integral is over the future-pointing part of the mass shell $ p^\mu p^\nu g_{\mu \nu} = -m^2$ making $p_0$ and $p^0$ functions of the integration variables $p^i$, the square brackets enclose an invariant measure on this space, and $f$ is a scalar distribution function on one-particle phase space (the mass shell within the tangent bundle) which we do not need to specify further here~\citep{ehlerssachs1968}.

\subsection{Weak-field, slow-motion expansion scheme}

We can develop an expansion scheme based on a nonrelativistic velocity $v \ll 1$ and anticipating that the potentials $\Psi$ and $\Phi$, in this section collectively denoted as $\phi$, behave similarly to Newtonian potentials and obey a Poisson equation sourced dominantly by the matter density fluctuations corresponding to the observed morphology. For simplicity and because we are considering the late universe we shall set $a \approx 1$ in the following explanation; comoving spatial coordinates $x$ can be thought of as physical.

 First, we make a weak-field assumption, neglecting neutron stars and black holes, and assume that the potentials are small of order $v^2 \ll 1$. Here $v$ plays the role of an escape velocity of a particle in the potential. It is numerically of order $v \sim 10^{-2}$. In the following, we will drop \emph{relative} corrections of order $v^2$ or higher, so e.g.~$\mathcal{O}(v^4)$ from the potentials.
 
 Second, spatial derivatives are large in the following sense. The linearity of the Poisson equation allows to decompose its solution into parts sourced by parts of the the hierarchy of structures of various sizes $L_i$ and locations $x_i$. Picking a part $\phi_i$ of the solution that dominantly contributes at some point $x$, 
 let us first assume $|x - x_i| \sim L_i$ (or closer). Since $\phi_i$ solves the Poisson equation, $\phi_i(x) = f_i((x-x_0)/L_i)$ where $f_i$ is well approximated by a low-order Laurent polynomial in the modulus of its argument $z \sim 1$ (or smaller) which ensures that $|d f_i(z) / dz |\sim |f_i| / z \sim |f_i| $ (or larger). Indeed, typical forms of $f_i(z)$ are $\propto 1/z$ outside of an overdensity or $\propto z^2$ in a region of homogeneous density (as below). By the chain rule, $|\nabla \phi_i| \sim \phi_i/L_i$. For each relevant term, its size $L_i$ naively is at most roughly the distance particles can have traveled in a Hubble time $H_0^{-1}$. The largest voids will then have a size of roughly $L_i \sim v/H_0$, but typical sizes of haloes, sheets and filaments (or even the solar system) can be much smaller. This, a spatial derivative with respect to any component of $x$ increases the value of $\phi$ by a factor of at least $L_i^{-1}$ per term, so by \emph{at least} $H_0/v$ overall,
 \begin{align}
 \nabla \gtrsim H_0 v^{-1}. \label{eq:spatiallarge}
 \end{align} 
 Therefore, $\nabla^2 \phi \sim H_0^2$ or larger; indeed we were anticipating $\nabla^2 \phi$ to be of order $G_N \rho \gtrsim G_N \rho_b \sim H_0^2$ when saying the potentials are similar to the Newtonian one. If $|x - x_i| \sim L_i$ does not hold for some large perturbation, which may contribute significantly even at a greater distance, we again assume that the size of voids is bounded from above and, further, that the largest ones are abundant enough that at least one is in the vicinity of every point. Then~(\ref{eq:spatiallarge}) still approximately holds.   
 
 Third, compared to spatial derivatives, time derivatives are suppressed in all cases by a factor of $v \ll 1$ if the potentials are changing due to the change of matter moving with such speed, leading to a quasi-static approximation. For simplicity we keep ``cosmological'' terms like 
 \begin{align}
 \label{eq:exprate}
H_a^2 := (\dot{a} / a)^2
 \end{align}  in the equation on all scales, even though on small scales where $\nabla^2 \phi \gg H_0^2$ and thus $\ddot{\phi} \gg H_0^2 v^2 $, it is possible that $\ddot{\phi}$ (which we drop in comparison to $\nabla^2 \phi$) is larger than $H_a^2$ (which for the standard scale factor $a$ is equal to $H_0^2$ today). Proceeding in this way is still consistent, because in this case the cosmological terms similar to~(\ref{eq:exprate}) are anyway strongly dominated by the second spatial derivatives $\nabla^2 \phi$ by a factor of more than $v^{-2}$. These cosmological terms therefore and do not spoil the result, so removing them from the equations on very small scales is not necessary. 
 
 A convenient consequence of this expansion scheme to leading order is that the equations become linear. A term quadratic in the potentials starts out at $ \mathcal{O}(v^4)$ and ends up at most at $\mathcal{O}( v^2)$ after two spatial derivatives. Higher-order derivatives do not appear in the Einstein equations.

 We assume further that the vector perturbation $\mathbf{B}$ is of order $\mathcal{O}(v^3)$ and behaves similarly to the potentials under derivatives. For further discussion of the scheme and the literature it is more practical to see the equations first.

\subsection{Field equations}

In the field equations in this section, we drop relative corrections of order $\mathcal{O}(v^2)$.
Given the expansion scheme, we obtain a Poisson equation 
\begin{align}
\label{eq:poisson}
3 H_a^2 + 2 a^{-2} \nabla^2 \Phi - \Lambda  = 8 \pi G_N T_{00} 
\end{align}
from the time-time-component, validating the assumptions we have been making on $\Phi$ in the previous subsection for developing the expansion scheme.\footnote{It should be noted at this point that seemingly circular reasoning like this is not problematic when solving differential equations as long as the final solution is self-consistent and satisfies all required boundary conditions under some assumptions ensuring uniqueness of the solution. (We also assume that an \emph{exact} solution is obtainable continuously from minute deformations of the approximate result.)  Instead of circular reasoning, this approach may be seen as an educated guess of a solution. There are however subtleties: besides not discussing tensor perturbations here, boundary conditions, required for uniqueness, will be addressed below. For now, we are still in the process of checking self-consistency.}  

 Further we get
\begin{align}
(i\neq j): \qquad\qquad \partial_i \partial_j (\Psi-\Phi)  &= 0  \label{eq:mixed} \\
(\partial_i^2 - \partial_j^2) (\Psi - \Phi)  &= 0, \label{eq:mixed2} \\
- 3 H_a^2 - 6 \frac{\ddot{a}}{a} + 3 \Lambda + 2a^{-2}\nabla^2(\Psi - \Phi) &= 0 \label{eq:mixed3} 
\end{align}
with the first set of equations coming from the spatial off-diagonal Einstein equations and the rest from combining the spatial diagonal equations. The energy-momentum tensor can be neglected here if it is of order $\rho v^2$ which would only give $\sim v^4$ corrections for the potentials after integrating the Poisson equation. 

This requires that \emph{either} the coordinates are approximately adapted to the Hubble flow that emerged from the homogeneous history of the universe so that 
\begin{align}
	|\langle \boldsymbol{v} \rangle|^2  \ll 1 \label{eq:adaptionHubbleFlow}
\end{align}
where the components of $\boldsymbol{v}$ are $v^i = T^{0i}/T^{00}$ and where the average is over a statistical ensemble at any location,
\emph{or} we are in the subhorizon regime where any Hubble flow itself is small,  
\begin{align}
|H_a \boldsymbol{x}|^2 \lesssim v^2 \ll 1, \label{eq:subhorizonregime} \\ 
H_a \sim H_0. \nonumber 
\end{align}
Note that in the second case, $\mathcal{O}(H_0)$ departures of $H_a$ are allowed. 

These equations determine the potentials. We check consistency with the space-time components below. 
The solution of~(\ref{eq:mixed}) is
\begin{align}
\Psi = \Phi + f_1(x,t) + f_2(y,t) + f_3(z,t). \label{eq:psi}
\end{align}
\changedagain{Most authors set the functions $f_i = 0$,\footnote{It is also easy to miss these terms when working in Fourier space.} but staying general will allow us to understand the  gauge freedom of the scale factor itself on small scales. Indeed, equations~(\ref{eq:mixed2}) only imply for the form of the functions $f_i$ evaluated at a point $(q,t)$} 
\begin{align}
f_i(q, t) = \alpha(t) q^2/2 + \beta_i(t) q + \gamma_i(t), \label{eq:fi}
\end{align}
 and thus~(\ref{eq:mixed3}) becomes
\begin{align}
2 \alpha(t) = a^2\left(H_a^2 +  2\frac{\ddot{a}}{a} - \Lambda  \right)  \equiv - k(t)  \label{eq:aandalpha}
\end{align}
so that~(\ref{eq:psi}) is with~(\ref{eq:fi}) 
\begin{align}
\label{eq:psiminusphi}
\Psi - \Phi = -k(t)  \frac{x^2 + y^2 + z^2 }{4} + \gamma(t) ,
\end{align}
where $\gamma(t) = \sum_1^3 \gamma_i(t)$ and where we have dropped a linear part $\bm{\beta}(t) \cdot \boldsymbol{x} $ as discussed in Section~\ref{sec:inertialframes}. We now also set $\gamma(t) = 0$. \changedagain{From comparison of~(\ref{eq:aandalpha}) with the usual Friedmann equations derived on the homogeneous and isotropic background level including spatial curvature explicitly, the term $k(t)=-2\alpha(t)$, which appears also in~(\ref{eq:psiminusphi}), admits an interpretation as time-dependent spatial curvature. Using~(\ref{eq:aandalpha}), we can chose either $a(t)$ and determine $k(t)$ by differentiation or choose $k(t)$ and determine $a(t)$ by integration. This works as long as} \emph{either}
\begin{align}
|k(t)| \ll H_0^2,
\end{align}
\emph{or} in the subhorizon regime~(\ref{eq:subhorizonregime}),  
since then the difference~(\ref{eq:psiminusphi}) of the two potentials is much smaller than unity.  

Reiterating on the comment around equation~(\ref{eq:exprate}) above, on \emph{very} small and fast scales, the result~(\ref{eq:psiminusphi}) should be thought of as $\Psi-\Phi \sim 0$ (to lowest order). The right hand side is a cosmological 
term $\propto k(t) r^2$ that automatically loses significance on small scales. In fact, near $r=0$ the Laplacian of $\Psi-\Phi$ may actually be dominated by space-dependent post-Newtonian terms. These new terms arise from solving equations to next order containing now both terms like $\ddot{\phi}$ in~(\ref{eq:mixed3})  as well as quadratic terms like $\phi \partial^2 \phi$ in~(\ref{eq:mixed}) (which are of similar size). 
However, since $\partial^2 \phi$ still dominates $\phi \partial^2 \phi$, the post-Newtonian terms are sub-leading and the statement $\Psi = \Phi$ (up to a constant) is now correct to leading order if the cosmological term $\propto k(t) r^2$ is even smaller. This is for example true when restricting to solar system scale, where we know how small the post-Newtonian corrections are. Therefore, while one could drop the cosmological term in e.g.~equation (\ref{eq:mixed3}) on such scales to emphasize it is not the largest subleading correction, it is not \emph{necessary} when the goal is only a valid leading-order treatment. As written, the equation holds also on cosmological scales. 

On the other hand, studying the full linearized Einstein equations including terms like $\ddot{\phi}$ in~(\ref{eq:mixed3}) while setting $\Psi = \Phi$, which removes spatial derivatives from~(\ref{eq:mixed3}) (and which is obtained from excluding terms like $\phi \partial^2 \phi$ and gauging $\alpha(t) =0 $) leads to unphysical results for large density contrasts~\citep{rasanen2010FLRW} and is incorrect given the physical setup because $\ddot{\phi} \sim \phi \partial^2 \phi$. Blind application of the standard cosmological perturbation theory treatment developed for small density contrasts, when a perturbative expansion in powers of the potential is well-behaved, does break down for large scale structure. This does not imply that proper leading-order physics cannot be obtained from a linearized treatment like ours. Instead of being forced to include quadratic terms or higher, the easier solution is to \emph{drop certain linear} terms by changing to an expansion scheme that tracks powers of $v$.

The expansion scheme discussed here can also effectively be obtained in Fourier space of linear perturbation theory in a subhorizon limit, which we see a posteriori. 
By taking the leading terms when $k \rightarrow \infty$ and before any manipulations that use $\delta \ll 1$ (and by selecting a Newtonian source) the expansion discussed here is recovered, but only for the special gauge of $\alpha(t) =0$. 
In the Fourier space approach polynomial growth in real space is easily missed. Typically, one introduces background equations to define sources of vanishing volume average to be able to divide by $k^2$.

\subsection{Friedmann equations as a gauge choice}
\label{sec:friedmannasgauge}

Comparing~(\ref{eq:psiminusphi}) and~(\ref{eq:aandalpha}) it is immediate that the standard choice $\Psi = \Phi$ implies $\alpha(t)=0$ (and $\gamma(t) = 0$). On small scales, this is a gauge choice. 
Vanishing $\alpha(t)$ in~(\ref{eq:aandalpha}) together with~(\ref{eq:poisson}) means that $a(t)$ satisfies both standard Friedmann equations for pressureless matter
\begin{align}
3 H_a^2 &= 8 \pi G_N \rho_b + \Lambda  \label{eq:fried1} \\
6 \frac{\ddot{a}}{a} &= - 8\pi G_N \rho_b + 2 \Lambda  \label{eq:fried2}
\end{align} if $\Phi = 0$ and $T_{00} = \rho_b$ is the background density, e.g.~on average. 
However, $\alpha(t) = 0$ in~(\ref{eq:aandalpha}) alone already implies the sum of the  two previous equations~(\ref{eq:fried1}) and~(\ref{eq:fried2})
\begin{align}
H_a^2 +  2\frac{\ddot{a}}{a} = \Lambda \label{eq:Friedmann3}
\end{align} \emph{without} invoking an averaging argument in the pressureless case considered here. The substitution $a(t) = h^{2/3}(t)$ reduces this to an inverted harmonic oscillator equation for $h$. The solutions admitting a big bang (with a negative sign in front of the decaying exponential) are then
\begin{align}
a(t) &= A \sinh \left(\frac{\sqrt{3 \Lambda}}{2} (t-t_0) + s \right)^{2/3}, \label{eq:matterlambda} 
\end{align}
where the integration constants $A, s$ can be fixed to have $a(t_0) = 1$ and $H_a(t_0) = H_0$ as 
\begin{align}
A &= (3 H_0^2 \Lambda^{-1} - 1)^{1/3} \equiv (\Omega_\Lambda^{-1} -1)^{1/3}  \\ 
s &=  \tanh^{-1}\left( \sqrt{ \Lambda (3 H_0^2)^{-1}} \right)\equiv \tanh^{-1}\left( \sqrt{ \Omega_\Lambda} \right),
\end{align}
which is already equivalent to the solution of the Friedmann equation~(\ref{eq:fried1}) for pressureless matter $\rho_b \sim a^{-3}$. \changedagain{Note that for fixed $\Lambda$ the normalization of the matter density enters indirectly into the integration constants via $\Omega_\Lambda = \Lambda / 3 H_0^2 = (8\pi G_N \rho_{b,0}/\Lambda + 1)^{-1}$, even if the differential equation did not contain a matter term. 
The $a^{-3}$ scaling of $\rho_b$ can be recovered from the solution~(\ref{eq:matterlambda}) together with~(\ref{eq:fried1}) (which was obtained from~(\ref{eq:poisson}) on average).  }

From~(\ref{eq:psiminusphi}), any other gauge choice implies the set-in of a quadratic divergence for the difference of the two potentials, and therefore for at least one them in the metric.
 The Friedmann gauge choice is thus relevant when constructing a global perturbed FLRW solution, to be discussed below in Section~\ref{sec:emergenceofa}.

If we are interested in a patch of the universe of size much smaller than the Hubble scale, a growth of the potentials $\sim H_0^2 r^2$ is of no concern. In this case $a(t)$ is allowed to depart from the Friedmann solution by $\mathcal{O}(1)$.
 Instead of satisfying~(\ref{eq:Friedmann3}), $a(t)$ may satisfy~(\ref{eq:aandalpha}) for $\alpha(t)\neq 0$ chosen with $\alpha(t) \sim H_0^2$. This preserves the condition for small peculiar velocities on subhorizon scales~(\ref{eq:subhorizonregime}). The scale factor is therefore directly related to an arbitrary gauge freedom in the subhorizon limit. 

As discussed above, $\Psi - \Phi$ also contains small nonlinear terms and time derivatives of the potentials, which dominate the quadratic difference on sufficiently small scales, so $\Psi\simeq\Phi$ restricted to very small scales does not imply Friedmannian scale factor evolution. Instead, the scale factor remains unconstrained in its cosmological evolution. This is expected, because the choice of scale factor should become completely irrelevant on these local scales. In particular it may still be chosen to obey a gauge like~(\ref{eq:matterlambda}) for a treatment unified with larger scales. 

\subsection{Motion of matter}
\label{sec:matter}

We now show that we indeed obtain the proper Newtonian limit independent of the choice of scale factor. 

First, the remaining time-space component of the field equations gives
\begin{align}
2 H_a \nabla \Psi+ 2 \partial_t \nabla \Phi +\frac{1}{2 a^2}\nabla \times (\nabla  
\times \mathbf{B})  =  8 \pi G_N T_{0i}. \label{eq:momentumconstraint}
\end{align}
Note that without a vector perturbation $\mathbf{B}$ of the metric, the current $T_{0i}$ would have to be longitudinal, which is not compatible with cosmological dynamics: neglecting the vorticity (small before stream crossing~\citep{jelic2018vorticity}), $\nabla \times (\rho \mathbf{v}) \approx (\nabla \rho) \times  \mathbf{v}$ and there is no reason for density (not potential!) gradient and velocity to be aligned. 
We may choose the vector potential of the transverse part of $16 \pi G_N a^2 T_{0i}$ at every instant to be transverse itself, and can set $ \mathbf{B}$ to be the vector potential of this quantity. Thus, $\nabla \times (\nabla 
\times \mathbf{B})/2 a^2 = 8 \pi G_N T_{0i}^T$, and we have absorbed the transverse part of $T_{0i}$. $\mathbf{B}$ may be chosen transverse as well (like in the Poisson gauge~\citep{bertschinger1995notes}). Then the double-curl of $\mathbf{B}$ equals the vector Laplacian of $\mathbf{B}$.\footnote{It appears that in this form the vector modes of FLRW were already discussed in~\citep{thomas2015vectors}.} Since the Laplacian converts the other potentials from $\mathcal{O}(v^2)$ to $\sim \rho$, when we assume the same type of spatial variation for $\mathbf{B}$ it is indeed smaller of order $\mathcal{O}(v^3)$ such that we have $\nabla^2 \mathbf{B}  \sim T_{0i} \sim \rho v$. Subtracting the transverse part, we are left with 
\begin{align}
2 H_a \partial_i \Psi+2 \partial_t \partial_i \Phi =  8 \pi G_N T_{0i}
\end{align}
where the source can be assumed to be longitudinal. 
We therefore do not lose information when taking the divergence. We get with~(\ref{eq:psiminusphi})
\begin{align}
(H_a +\partial_t) 2 \nabla^2 \Phi  + 3 a^2 H_a \left(H_a^2 +  2\frac{\ddot{a}}{a} - \Lambda \right)=  8 \pi G_N\partial_i T_{0i}.
\end{align}
Using the Poisson equation~(\ref{eq:poisson}) we can replace $2 \nabla^2 \Phi = 8 \pi G_N a^2 T_{00} - 3 a^2 H_a^2 + a^2 \Lambda$, and it can be directly verified that many terms cancel to give the equation $\partial_t T_{00} + 3 H_a T_{00} = a^{-2}  \partial_i T_{0i} $, or, dropping terms of order $\mathcal{O}(\partial_t T_{00} v^2)$ and $\mathcal{O}(\partial_i T_{00} v^3)$ with respect to $\mathcal{O}(\partial_t T_{00})$, that is, up to relative corrections of $\mathcal{O}(v^2)$ which we must drop,  
\begin{align}
\partial_t T^{00} + 3 H_a T^{00} = - \partial_i T^{0i}
\end{align}
where we have raised indices while the metric is still approximately diagonal. Going to physical \changedagain{(meaning not comoving)} spatial coordinates with $r^i = a x^i$ and $t$ unchanged does not change $T^{00}$, but $T^{0i} \rightarrow - H_a x^i T^{00} + T^{0i}/a$ (using the same name for the tensor components in the new coordinates). Since $\partial_i \rightarrow a \partial_i$ this gives simply the Newtonian continuity equation
\begin{align} 
\label{eq:newtoncontinuity}
\partial_t T^{00}  + \partial_i T^{0i} = 0.
\end{align}

To obtain a leading-order equation of motion for the matter it is easiest to expand the geodesic equation directly to required order. With the four-velocity $u^\mu = d x^\mu / d\tau$ it can be written without Christoffel symbols as 
\begin{align}
\frac{d u_\sigma }{d \tau} = \frac{1}{2} \partial_\sigma g_{\mu\nu} u^\mu u^\nu. \label{eq:geodesic}
\end{align}
We are back in the comoving coordinates with scale factor $a$. With $u^\mu = (-1, \partial_t x^i ) + \mathcal{O}(v^2)$ we have $u_i = a^2  \partial_t x^i + \mathcal{O}(v^3)$. 
The left hand side of~(\ref{eq:geodesic}) with $\sigma = i$ can be expressed with the physical coordinates $\boldsymbol{r} = a \boldsymbol{x}$ as $\partial_t (a^2 \partial_t \boldsymbol{x}) = a \partial_t^2 \boldsymbol{r} - \boldsymbol{r} \partial_t^2 a$. 
The right hand side is $(1+\mathcal{O}(v^2)) \nabla g_{00}/2  \approx  - \nabla \Psi \rightarrow -a \nabla \Psi $ where the last gradient is with respect to $\boldsymbol{r}$. 
This means that in physical coordinates
\begin{align}
\label{eq:accel}
\ddot{\boldsymbol{r}}  = - \nabla \Psi + \frac{\ddot{a}}{a} \boldsymbol{r}.  
\end{align}
We can compare this to the standard Newtonian case involving the gradient of a potential sourced by the \emph{total} matter content as 
\begin{align}
\nabla^2 \phi_N = 4 \pi G_N T_{00}. \label{eq:poissonnewton}
\end{align}
Comparison of the two Poisson equations (\ref{eq:poissonnewton}) and (\ref{eq:poisson}) (absorbing the $a^2$ in the latter into the new Laplacian with respect to physical coordinates and assuming adapted boundary conditions) gives 
\begin{align}
\Phi &= \phi_N + (\Lambda - 3 H_a^2)  \boldsymbol{r}^2/12 \label{eq:phiphiN} \\ 
-\nabla \Phi &= -\nabla \phi_N + (H_a^2 /2 - \Lambda /6 ) \boldsymbol{r}.
\end{align}
 We can trade the $\Psi$ potential in~(\ref{eq:accel}) for the $\Phi$ potential with~(\ref{eq:psiminusphi}) and finally for the Newton potential $\phi_N$  
\begin{align}
\ddot{\boldsymbol{r}}  &= - \nabla \Psi + \frac{\ddot{a}}{a} \boldsymbol{r} \\
& = - \nabla \Phi -   \frac{1}{2} H_a^2  \boldsymbol{r} + \frac{1}{2} \Lambda \boldsymbol{r}  \label{eq:fullforce} \\ 
& = - \nabla \phi_N + \frac{1}{3} \Lambda \boldsymbol{r} . \label{eq:NewtonianGeodesic}
\end{align} 
We stress that the result is that for $\Lambda = 0$ there is no cosmic force on small scales at all, consistent with the starting point of Newtonian cosmology.

\section{Embedded top-hat collapse} 
\label{sec:tophat}
To test the gauge freedom of the scale factor and the applicability of the perturbed FLRW metric in situations of strong inhomogeneity, we consider a spherically symmetric example with $\Lambda = 0$ for simplicity. For radial motion, $T_{0i}$ is irrotational and we can set $\mathbf{B}=0$ in the following. In particular, we consider the collapse of a pressureless dust top-hat (core region), leaving behind a vacuum shell, embedded into the matter cosmos.  

To understand our example it is useful to be aware of the Einstein--Straus solution~\citep{einsteinstraus1945, einsteinstraus1946, carrera2010influence}. It serves as an even simpler toy model of a collapsed structure in the expanding universe and can be thought of as the end result of a spherical collapse in the following sense: it represents what is obtained from a matter-dominated universe by compressing the dust contained within a ball of arbitrary size into a black hole. The dust outside of the vacuum bubble created in this way continues its usual expansion not only in Newtonian gravity but also, exactly, in GR. The vacuum bubble is therefore bounded by a shell of fixed comoving radial coordinate  and expanding in physical coordinates (for standard comoving coordinates adjusted to the initial Hubble flow). An observer entering the bubble from the outside trying to remain at fixed comoving \emph{or} physical coordinates needs to compensate a gravitational pull towards the center that is increasing with decreasing distance to the black hole (starting at zero for the comoving case) and ultimately diverging at the Schwarzschild horizon, but the crossover into the bubble is smooth. An explicit solution in a single coordinate system is rather unwieldy in terms of formulas~\citep{schuecking1954, laarakkers2001einsteinstraus, balbinot1988einsteinstraus}, considerably more so than the solution McVittie found by direct integration of the Einstein equations~\citep{mcvittie1933, nandra2012mcvittie} with a similar goal of describing a point mass embedded in an expanding universe, but the physical picture is simpler for the Einstein--Straus solution. 

\begin{figure}
	\includegraphics[width=1.0\columnwidth]{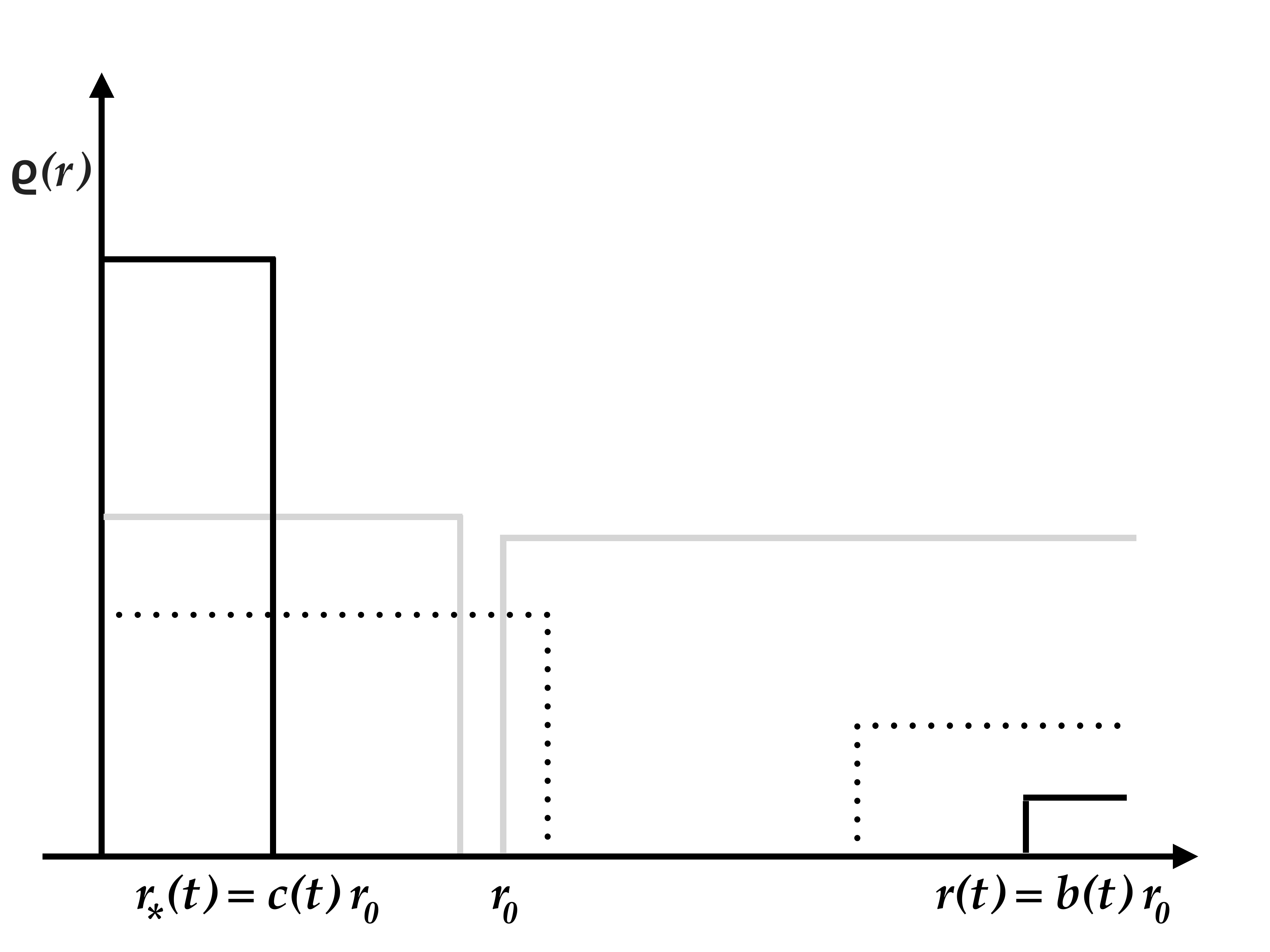}
	\caption{An example initial density profile is shown (grey) with a top-hat created by compressing the matter in a homogeneous universe within a spherical region. After some expansion (dotted), the top-hat turns around and collapses whereas the outer universe continues to expand (black). The boundary of the top-hat is denoted as $r_*(t)$ and the outer boundary of the vacuum region as $r(t)$. Their time evolution is parametrized by two scale factors $c(t)$ and $b(t)$, respectively. }
	\label{fig:density}
\end{figure}

Instead of imposing a central black hole, we will be concerned with the previous weak-field stage of the collapse, for which the black hole is smoothed to a collapsing top-hat core similar to the Oppenheimer--Snyder model~\citep{oppenheimersnyder1939}. The Oppenheimer--Snyder model is in some sense the most extreme yet simple case of collapse. The exact solution of a top-hat undergoing turnaround and collapse is a portion of the closed FLRW spacetime~\citep{MTW2017collapsestar}\footnote{The original interior Oppenheimer--Snyder solution~\citep{oppenheimersnyder1939} is actually equal to a patch of time-reversed flat EdS which does not allow for a turnaround.}. The combined Einstein--Straus--Oppenheimer--Snyder collapse has been discussed by other authors within GR~\citep{nottale1982tophat, dai2015separateuniverses, kim2018spherically}.  The density profile and its evolution is schematically illustrated in Figure~\ref{fig:density}.

 We wish to model this spacetime in the perturbed FLRW gauge, which will give a continuously differentiable but approximate solution in a single coordinate system by construction. Our strategy is to assess the quality of the approximation by comparing the result to the exact (closed FLRW, Schwarzschild, flat FLRW) solutions separately for each of the three regions by finding appropriate coordinate transformations. This compliments the more complicated and less explicit result in~\citep{karel2008LTBFLRW}, which studies the validity of the perturbed FLRW solution by comparison to the exact LTB class of solutions, with the explicit top-hat collapse case that intuitively is the most challenging one for the perturbed FLRW metric.\footnote{A reader familiar with the LTB solutions may think that the embedded top-hat collapse should be a special case. Indeed, LTB solutions exactly describe more general inhomogeneous dust distributions with spherical symmetry. However, a single LTB metric is not able to cover the spacetime considered with one set of coordinates. The gauge ($g_{tt}=-1$, diagonal) is too constraining for each region so that one inherits the discontinuities from the discontinuous source.}

\subsection{Newtonian top-hat collapse}
\label{sec:tophateqns}

We first use a physical radial coordinate $r$ in which Newtonian theory holds in the usual way to describe some aspects of the collapse. Subsequently, we will work with two alternative choices of comoving coordinates.\footnote{To be clear, again the word ``comoving'' refers only to the use of FLRW coordinates with some scale factor of the metric~(\ref{eq:lineelement}) factored out from the physical spatial coordinates. Comoving coordinates in this sense do not guarantee vanishing or small peculiar velocities, because the scale factor may not be related to the physical motion of matter.}
 While it would be possible to stay entirely within the respective comoving coordinates for the calculations, this would double part of our work. 

In physical coordinates, we can use Newtonian gravity according to Section~(\ref{sec:matter}).
We let the initial radius of the ball be $r_0$, so the conserved mass in the top-hat is 
\begin{align}
M = \frac{4 \pi}{3}  r^3_0 \rho_{b,0} 
\end{align} where $\rho_{b,0}$ is the initial cosmic background density.\footnote{Conservation of mass is implied by the continuity equation~(\ref{eq:newtoncontinuity}). The flat-space expression for the volume of the ball follows only to leading order. Metric perturbations and perturbations of $T_{00}$ and of the time coordinate that would change the spatial slicing due to coordinate transformations relevant in the present work can be neglected at the required order, because the mass only enters into small terms that are themselves perturbative. }
The expansion of the boundary of the ball can be parametrized as $r(t) = b(t) r_0$ with $b(t_0) = 1$. Then $b(t)$ obeys
\begin{align}
H_b^2 = \frac{\dot{b}^2}{b^2} = \frac{2 G_N M}{r_0^3 b^3},  \label{eq:scalefactorb}
\end{align} 
which follows from setting the conserved total energy\footnote{The energy conservation employed here is the first integral of the Newtonian form of the geodesics (\ref{eq:NewtonianGeodesic}).} for the shell to zero so that $\dot{r}^2(t) /2 = E_{kin} = - E_{pot} = G_NM/r(t)$, assuming shells do not cross so that $M=M(r_0) \sim r_0^3 $ is constant in time. No crossing is ensured if the shells further out at $r_0' > r_0$ expand with the same law $r'(t) = b(t) r_0'$, or equivalently, if their total energy is also vanishing, which we assume as an initial condition.\footnote{One may also check that the continuity equation preserves homogeneity of the density under the linear radial velocity profiles that the law $r(t) = b(t) r_0$ implies at an initial time, which in turn are preserved by the gravitational force of a homogeneous ball. Of course, this is instable: gravitation amplifies small inhomogeneities. A real top-hat disintegrates into shocks; the present model is a smoothed idealization.}

Equation~(\ref{eq:scalefactorb}) is equivalent to the Friedmann equation for the flat matter-dominated Einstein--de Sitter (EdS) universe with a Hubble parameter $H_0 \equiv H_b(t_0)$ at the initial time\footnote{Note that $t_0$ may be chosen to be e.g.~in the past and therefore $H_0$ may not mean the Hubble parameter today.} obeying 
\begin{align}
H_0^2 r_0^3 = 2 G_N M =  \frac{8 \pi G_N}{3} r^3_0 \rho_{b,0},
\end{align} so that 
\begin{align}
H_b^2 = \frac{H_0^2}{b^3} =  \frac{8 \pi G_N}{3} \rho_{b} \quad \text{where} \;\; \rho_b \equiv b^{-3} \rho_{b,0} \label{eq:scalefactorb2}
\end{align} 
that is, to~(\ref{eq:fried1}) with vanishing cosmological constant, and therefore to~(\ref{eq:Friedmann3}), under the renaming $b(t) \rightarrow a(t)$. However, these equations a priori do not describe the same thing, since $b(t)$ is related to the physical motion of matter shells in the outer region whereas $a(t)$ is the particular metric scale factor choice for which the potentials do not diverge.  
However, if horizon scales are to be considered, we saw around equation~(\ref{eq:adaptionHubbleFlow}) that the Hubble flow must be aligned to the coordinates to ensure small peculiar velocities, so the equations for $a(t)$ and $b(t)$ must approximately coincide. That is why we have chosen vanishing total shell energy at large distances here. Otherwise, there would be an additional term $K/b^2$ in~(\ref{eq:scalefactorb2}), corresponding to a spatial curvature term in a Friedmann equation, which would only be consistent with a non-flat metric ansatz at large scales.   


Locally, however, the Newtonian theory coming from the flat perturbed FLRW metric does allow for the set-up of the initial state corresponding to other value of the shell energies. 
In particular, we assume that the top-hat is much smaller than the Horizon size. The top-hat collapse evolution is independent of the dynamics outside and can be parametrized by another homogeneous scale factor $c(t)$ with $r_*(t) = c(t) r_0$ where $r_*(t)$ is the radial position of the out-most shell at time $t$. Choosing a negative total energy $E$ for all top-hat shells we easily find  
\begin{align}
\label{eq:scalefactorc}
H_c^2 =\frac{\dot{c}^2}{c^2} = \frac{H_0^2}{c^3} + \frac{K}{c^2}
\end{align}
 where $K = 2 E / r_0^2 < 0$ by choice. Here the definition of $H_0$ is unchanged from before in terms of the mass $M$; $H_0$ is not necessarily to be understood as the value of $H_c$ at $t_0$ (but still of $H_b$, as before). The actual value of $H_c(t_0)$ depends on $K$ and $c(t_0)$. We could, for example, choose $H_c(t_0) = H_0$ by setting $c(t_0) < 1 $ appropriately for a given $K$, corresponding to an initial state of a compressed top-hat that velocity-wise is initially aligned with the previous Hubble flow.  At later times, the motion of the matter described by $c(t)$ does not correspond to the (parabolic) expansion law $b(t) \propto t^{2/3}$ of the matter outside. Indeed, eventually $c(t)$ will be a decreasing function with $H_c < 0$. We are not going to need the well-known exact (cycloidal) solutions to this Friedmann equation. 

We now first explore the natural choice $a(t) = b(t)$ and then the choice $a(t) = c(t)$ for the metric scale factor. The latter choice can be valid if not only the top-hat but the whole patch considered are small compared to the Hubble distance $H_a^{-1}$ according to the discussion after equation~(\ref{eq:adaptionHubbleFlow}).

\subsection{Cosmological scale factor}
\label{sec:cosmoscalefactor}

First, we make the standard choice $\Psi = \Phi \equiv \phi$ so that the metric scale factor is~(\ref{eq:matterlambda}) in the $\Lambda \rightarrow 0 $ limit, which gives the EdS solution. That is, we chose $a(t) = b(t)$. The metric~(\ref{eq:lineelement}) becomes
\begin{align}
\label{eq:metricb}
ds^2 = -(1+2\phi) dt^2 + b^2(t) (1-2\phi) (dx^2  + x^2 d\Omega^2)
\end{align}
where $x = b^{-1}(t) r$ is the comoving radius, 
\begin{align} 
b(t) = (3 H_0 t / 2)^{2/3}
\end{align}
  satisfies the Friedmann equation~(\ref{eq:scalefactorb2}) and the Poisson equation~(\ref{eq:poisson}) reads
\begin{align}
\nabla^2 \phi = 4 \pi G_N b^2  (\rho - \rho_b). \label{eq:poissonnormal}
\end{align}

\begin{figure}
	\includegraphics[width=1.0\columnwidth]{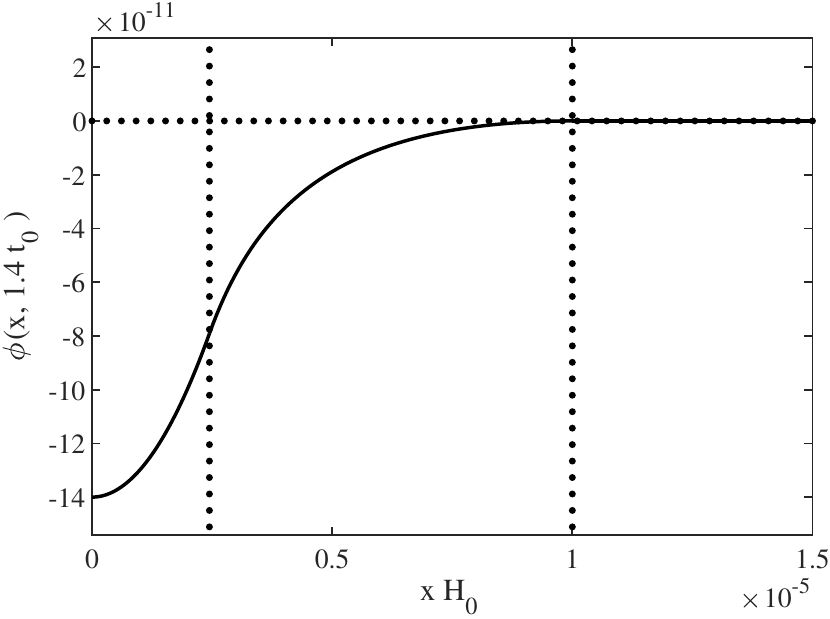}
	\caption{\changed{This plot shows the three matched pieces of the solution for $\phi$ in equations~(\ref{eq:inphi1},\ref{eq:vacuumphi1},\ref{eq:outphi1}) for an overdensity obtained from first compressing the dust in a ball of radius $r_0 = 10^{-5} H_0^{-1}$ to $0.4 r_0$ ($c(t_0) = 0.4$) and then letting it collapse until $t_f = 1.4 t_0$. The vertical lines indicate the separation into core, vacuum and outer cosmological region. The radial coordinate $x$ is comoving with scale factor $b(t)$. In the moment $t=t_f$ shown in the plot, $b(t_{f}) \approx 1.66$ (thus, if $t_{f}$ refers to today, $t_0$ is at a redshift of $0.66$).  Initial velocities were aligned with the original Hubble flow; it follows that $K=-2.34$. The density contrast in the top-hat core, in the leftmost area of the plot, is $\delta = b^3(t_f)/c^3(t_f)-1 \approx 67$, but the potential remains small everywhere. }}
	\label{fig:phi}
\end{figure}

We can solve this Poisson equation ensuring continuity of $\phi$ and $\partial_x \phi$ at the two boundaries starting from the inside. To reduce clutter, in the following units are such that $G_N = 1$. We obtain 
\begin{numcases}{\phi(x,t)=}
 \label{eq:inphi1} \frac{1}{b(t)} \left(  \frac{M}{2} \frac{ (x / x_*(t))^2  -3}{x_*(t)} -\frac{1}{4} H_0^2 \left( x^2 - 3  x_0^2\right)  \right)  \\ 
\label{eq:vacuumphi1} -\frac{1}{b(t)} \left( M/x + \frac{1}{4} H_0^2 \left( x^2 - 3  x_0^2\right) \right) \\  
\label{eq:outphi1} 0 
\end{numcases}
where the three formulas apply in the three regions $x \leq x_*(t)$, $x_*(t) < x < x_0$ and $x \geq  x_0$, respectively, and where $x_*(t) = b(t)^{-1} r_*(t) $ is the comoving position of the boundary of the top-hat and and $x_0 = r_0$ is the comoving position of the transition from vacuum to cosmological density. Here, $\phi$ has been shifted to zero outside. An exemplary $\phi$ is plotted in Figure~(\ref{fig:phi})

\subsubsection{External region $x \ge  x_0$}
\label{sec:external1}

Since $\phi = 0$, the metric~(\ref{eq:metricb}) is the FLRW metric for the EdS cosmos. 
\subsubsection{Vacuum region $ x_*(t) < x < x_0$ }
\label{sec:vacuum1}
We first transform~(\ref{eq:metricb}) to the areal radius $r = (1-\phi) b x$ (to be distinguished from the function $r(t) = b(t) r_0$), which yields, to first order in $\phi$,
\begin{align}
ds^2 &= &-(1+2\phi - (1-2\phi)A^2 B^{-2}) dt^2 \nonumber \\ 
& &- 2(1-2\phi)AB^{-2} dr dt \nonumber \\ 
& &+ (1-2\phi)B^{-2} dr^2 + r^2 d\Omega^2 \\
A &= &H_b r - \phi_{,t} b x \label{eq:A} \\
B &= &1 - \phi - \phi_{,x} x 
\end{align}
where the time derivative ${}_{,t}$ in~(\ref{eq:A}) is at constant $x$ and where afterwards $x$ should be eliminated in terms of $r$. This follows immediately from writing $dr = A dt +  b B dx \Leftrightarrow (b dx)^2 = B^{-2} (dr - A dt)^2$. 

In the following we keep $H_b^2 r^2$ and $M/r$ as the smallest terms. The first one is very small compared to one if the vacuum region is very small compared to the Hubble distance, $r_0 \ll H_b^{-1}$, because $x\le x_0$ and $b = \mathcal{O}(1)$; the second term is everywhere small under the weak-field assumption. We drop $\sqrt{H_b^2 r^2} (M/r) = H_b M$ and smaller, and thus also $H_b r \phi \ll v^2$. In general, one cannot drop terms of order $v^3$ (like $\mathbf{B}$ itself) from the vector mode; however, only the transverse part of any vector mode enters in the continuity equation~(\ref{eq:momentumconstraint}). We can therefore definitely ignore radial $r-t$ off-diagonal terms of order $v^3$ in the metric, corresponding to a longitudinal mode.

From the potential~(\ref{eq:vacuumphi1}), $A \approx H_br$. Also, $B \approx 1 + 3 H_b^2 (r^2 - r^2(t))/4$. 
This leads to 
\begin{align}
ds^2 = -&\left(1-2M/r - \frac{3}{2} H_b^2 (r^2 - r^2(t))\right) dt^2 - 2H_br dr dt \nonumber \\ + &\left(1+2M/r -H_b^2 r^2 \right) dr^2 + r^2 d\Omega^2. 
\end{align}
Our task is to remove the off-diagonal part by a transformation to a new time coordinate 
\begin{align} 
t' = f(t,r),
\end{align} since the radial coordinate is the areal radius already, just like in Schwarzschild coordinates. Let $J$ denote the Jacobian of the new coordinates as a function of the old ones; the metric transforms as $g \rightarrow (J^{-1})^t g J^{-1}$. Writing only the nontrivial $(t,r)$ part, we have 
\begin{align}
J = \begin{pmatrix} f_{,t} & f_{,r} \\ 0 & 1 \end{pmatrix}  \quad \rightarrow \quad J^{-1} = \frac{1}{f_{,t} }\begin{pmatrix} 1 & -f_{,r} \\ 0 & f_{,t} \end{pmatrix} 
\end{align}
and find 
\begin{align}
g  \quad \rightarrow \quad \frac{1}{f_{,t}^2 }\begin{pmatrix} g_{tt} & -f_{,r} g_{tt} + f_{,t} g_{rt} \\  -f_{,r} g_{tt} + f_{,t} g_{rt} & f_{,r}^2 g_{tt} - 2 f_{,r} f_{,t} g_{rt} + f_{,t}^2 g_{rr} \end{pmatrix} 
\end{align}
and can read off that for 
\begin{align}
\label{eq:diagPDE}
f_{,r} = \frac{g_{rt} }{g_{tt}} f_{,t} 
\end{align}
the metric transforms as
\begin{align}
g  \quad \rightarrow \quad \begin{pmatrix}  g_{tt} /  f_{,t}^2& 0 \\ 0 &  g_{rr}  -  g_{rt}^2/g_{tt}  \end{pmatrix} . \label{eq:newmetric}
\end{align}
The condition~(\ref{eq:diagPDE}) is a first-order linear PDE (that can be reduced to a nonlinear ODE by the method of characteristics). It may be noted that evidently it is not necessary to solve the PDE to get the new $g_{rr}$, which does not depend on $f$ in~(\ref{eq:newmetric}).  

In the present case, any diagonalizing transformation adds $-g_{rt}^2/ g_{tt}  \approx g_{rt}^2 = H^2 r^2 $ to $g_{rr}$ which therefore becomes $1+2M/r$. 
If the $t't'$-element was $-(1-2M/r)$, the metric would clearly be the weak-field limit of the Schwarzschild solution. 
All that is required for this is that 
\begin{align}
\label{eq:fcommat}
f_{,t} = 1 - \frac{\delta g_{tt}}{2}
\end{align}  where $\delta g_{tt} $ are the terms we want to remove, for then 
\begin{align} 
g_{tt} \quad &\rightarrow  \qquad \quad g_{tt} \qquad \qquad f_{,t}^{-2}  \nonumber \\ 
&\approx -\left(1-\frac{2M}{r} - \delta g_{tt} \right) (1+\delta  g_{tt} )\approx - \left(1-\frac{2M}{r}\right) , 
\end{align}
so $f_{,t} = 1 - 3 H^2 (r^2 - r^2(t))/4$. On the other hand, the PDE 
 is, to sufficient order and assuming $f_{,t} = O(H^2 r^2)$ (smallest possible transformation), simply $f_{,r}  \approx  g_{rt} / g_{tt} \approx  -g_{rt} =  H r$ so that $f = H r^2 /2 +  h(t)$. Given that $H = 2/(3t)$ and the form of $b(t)$ we see that there indeed exists a solution $f = t + H r^2 /2  + 3 (H_0 x_0)^2 / (2 H b)  $ with the required $f_{,t}$.
 
In summary, the coordinate transformation
\begin{align}
x' &\equiv r  = (1-\phi(x,t)) b(t) x \\ 
t' &= t + \frac{1}{2} H_b(t) b^2(t) x^2    + \frac{3 H_0^2 x_0^2}{2 H_b(t) b(t)}
\end{align}
puts the perturbed FLRW metric~(\ref{eq:metricb}) in the vacuum region~(\ref{eq:vacuumphi1}) into the weak-field approximation of the Schwarzschild solution. Note in particular that the metric is now static. 

\subsubsection{Interior region $  x \leq x_*(t) $}
\label{sec:internal1}

We can play a similar game for the interior region. 
We first introduce a spatial coordinate 
\begin{align}
\label{eq:xprime}
x' = (1-\phi - \epsilon )bx/c \approx (1-\phi)(1 - \epsilon) bx/c.
\end{align} Here $\epsilon = \epsilon(x,t)$ is assumed to be of the same order as $\phi$. When $\epsilon$ vanishes, this is a straightforward coordinate change designed taking $g_{\theta \theta}$ from $x^2 (1-2\phi) b^2(t)$ to $x'^2 c^2(t)$. The role of $\epsilon$ is to compensate the effect of a given further time transformation to a new time $t'$, such that instead $x^2 (1-2\phi) b^2(t) = x'^2 \tilde{c}^2(t')$ for a given choice of the function $\tilde{c}$. Thus, 
\begin{align}
\label{eq:epsbtilde}
(1+2\epsilon)  c^2(t) = \tilde{c}^2(t'). 
\end{align}

Again writing $dx' = A dt + b B dx$, we have now that 
\begin{align}
A &\approx (H_b - H_c) x' \nonumber \\ 
B &= c^{-1}(1- (\phi + \epsilon) - (\phi +\epsilon)_{,x} x) \label{eq:ABinterior}
\end{align} 
where we have anticipated that the rate of change $\dot{\epsilon}/\epsilon= \mathcal{O}(H_0)$ to neglect terms.

Using the Friedmann equation for $c$, (\ref{eq:scalefactorc}), the expression for $\phi$ in the interior region~(\ref{eq:inphi1}) can be written as 
\begin{align}
\phi &= \frac{1}{4} H_0^2 \left( b^2 x^2 (c^{-3} - b^{-3}) - 3 x_0^2 (c^{-1} - b^{-1})\right) \\ &= \frac{1}{4}\left[ (H_c^2 - H_b^2 ) c^2 x'^2 - K x'^2  + 3 H_0^2 x_0^2 (b^{-1} - c^{-1})   \right].
\end{align} 
With the definitions~(\ref{eq:ABinterior}) the metric~(\ref{eq:metricb}) becomes, under the transformation~(\ref{eq:xprime}),
\begin{align}
g_{tt} &\approx -(1+2 \phi - A^2 c^2) \nonumber \\
g_{x' t} &\approx - A c^2. \nonumber  \\ 
g_{x' x'} &= (1-2\phi ) B^{-2}  \label{eq:gspatialint}  
\end{align}
Again, for integrating~(\ref{eq:diagPDE}) (with $x'$ playing the role of $r$), the procedure reduces to integrating $-g_{x't }$ in $x'$, to give 
\begin{align}
\label{eq:f1}
 t' = f(x', t) = t + (H_b - H_c) c^2 x'^2 /2 + g(t).
\end{align}
 For $g_{t't'} \approx -1$ we would like again that~(\ref{eq:fcommat}) is satisfied, with $\delta g_{tt}$ now defined as all terms perturbing $-1$ in $g_{tt}$, which reads
\begin{align}
 f_{,t} = 1 + &\frac{c^2 (H_c^2 - H_b^2 - 2(H_c-H_b)^2) - K}{4}x'^2 \nonumber \\ + &\frac{3 H_0^2 x_0^2 (b^{-1} - c^{-1})} {4}. 
 \end{align}
 It can be shown, using the two Friedmann equations for $b$~(\ref{eq:scalefactorb2}) and $c$~(\ref{eq:scalefactorc}) to eliminate time derivatives, that this is indeed consistent with~(\ref{eq:f1}) when we pick 
 \begin{align}
g(t) = (3/4) H_0^2 x_0^2 \int   (b^{-1} - c^{-1}) dt . 
 \end{align}

The metric is now diagonal and $g_{tt} = -1$. We are ready to read off an $\epsilon$ after Taylor-expanding the left-hand side of~(\ref{eq:epsbtilde}) around $t'$, defining $\delta t = t' - t$,
\begin{align*}
c^2(t) &(1+2 \epsilon) \\
 &\approx \left([c^2]|_{t=t'} - [c^2]_{,t} \delta t\right)(1+2\epsilon) \\  
 &\approx c^2(t=t') + 2 c^2 \epsilon - c^4 x'^2 (H_c H_b - H_c^2) - 2 c^2 H_c g(t).
\end{align*}
This means that 
\begin{align}
\epsilon &= H_c g(t) + c^2 x'^2 (H_c H_b - H_c^2)/2 \nonumber \\ &\approx H_c g(t) + b^2 x^2 (H_c H_b - H_c^2)/2
\end{align} would imply that $\tilde{c}(t') = c(t=t')$ from~(\ref{eq:epsbtilde}). This is a convenient choice, for it means that $\tilde{c}$ has been chosen to satisfy the same Friedmann equation (with the time derivative with respect to the new time variable $t'$!) for the closed universe we have found earlier for $c$, because it is the same function of \emph{its} argument as is $c$ of \emph{its} argument, so this is the  natural choice we adopt. 

Finally, we compute $g_{x'x'}$, which picks up an additional $+ g_{x' t}^2$ compared to~(\ref{eq:gspatialint}) under the diagonalizing time transformation. Thus,
\begin{align}
g_{x' x'} &= (1-2\phi ) B^{-2}+ g_{x' t}^2 \nonumber \\
 &\approx c^2(t) (1+2(\phi + \epsilon)_{,x} x + 2 \epsilon ) + g_{x' t}^2 \nonumber \\
 &\approx \tilde{c}^2(t') (1+2(\phi + \epsilon)_{,x'} x')   + g_{x' t}^2 \nonumber\\
 &\approx c^2(t') (1+2(\phi + \epsilon)_{,x'} x' + c^2 x'^2  (H_c - H_b)^2 ).
\end{align} 
This simplifies straightforwardly to $g_{x' x'} = c^2(t') (1-K x'^2)$ for the choice of $\epsilon$ we have made. With the identification $k= -K > 0$ the metric is now 
\begin{align}
\label{eq:metrictophat}
 ds^2 = - dt'^2 + c^2(t')\left( \frac{dx'^2} {1-kx'^2} + x'^2 d\Omega^2\right)
\end{align}
for $0 < kx'^2 \ll 1$, which was already assumed. This is the metric of the closed universe as well as the exact solution of the collapsing top hat. 

In summary, the transformation
\begin{align}
x' & =  x(1-\phi(x,t))\left(1 - H_c(t) \delta t(x,t)\right) \frac{b(t)}{c(t)} \label{eq:trafo1} \\ 
t' &= t + \delta t(x,t)  \label{eq:trafo2} \\ 
\delta t(x,t) &\equiv \frac{H_b(t) - H_c(t)}{2} b^2(t) x^2 + g(t) \label{eq:trafo3} \\
g(t) &\equiv \frac{3}{4} H_0^2 x_0^2 \int   (b^{-1} - c^{-1}) dt  \label{eq:trafo4}
\end{align}
puts the interior region of the perturbed FLRW metric above into the weak-field closed FLRW form~(\ref{eq:metrictophat}) describing a collapsing top-hat dust profile.

\subsection{Collapsing scale factor}
\label{sec:collapsescalefactor}

We now choose $a(t) = c(t)$ as the metric scale factor, that is, coordinates that are collapsing with the top-hat. 
The metric ansatz~(\ref{eq:lineelement}) is then
\begin{align}
\label{eq:metricc}
ds^2 = -(1+2\Psi) dt^2 + c^2(t) (1-2\Phi) (dx^2  + x^2 d\Omega^2)
\end{align}
where $x = c^{-1}(t) r$ is the comoving radius, $c(t)$  satisfies the Friedmann equation~(\ref{eq:scalefactorc}) and the Poisson equation~(\ref{eq:poisson}) reads
\begin{align}
\nabla^2 \Phi = 4 \pi G_N c^2(t)  \rho(x,t)  - \frac{3}{2} c^2(t) H_c^2(t)  \label{eq:poissonc}.
\end{align}
We also need $\Psi$, which according to~(\ref{eq:psiminusphi}) is given by 
\begin{align} 
\Psi = \Phi + \frac{c^2}{2} \left( \frac{H_c^2}{2} + \frac{\ddot{c}}{c} \right) x^2 = \Phi + \frac{K}{4}x^2 
\end{align}
where the Friedmann equation for $c$ was differentiated in time to obtain the result. \changed{The interpretation of the prefactor of $x^2/4$ in the difference $\Phi-\Psi$ as a spatial curvature term $k(t)$ that was mentioned after equation~(\ref{eq:psiminusphi}) is of course recovered in this simple example, in which $k=-K$ is time-independent.}

\begin{figure}
	\includegraphics[width=1.0\columnwidth]{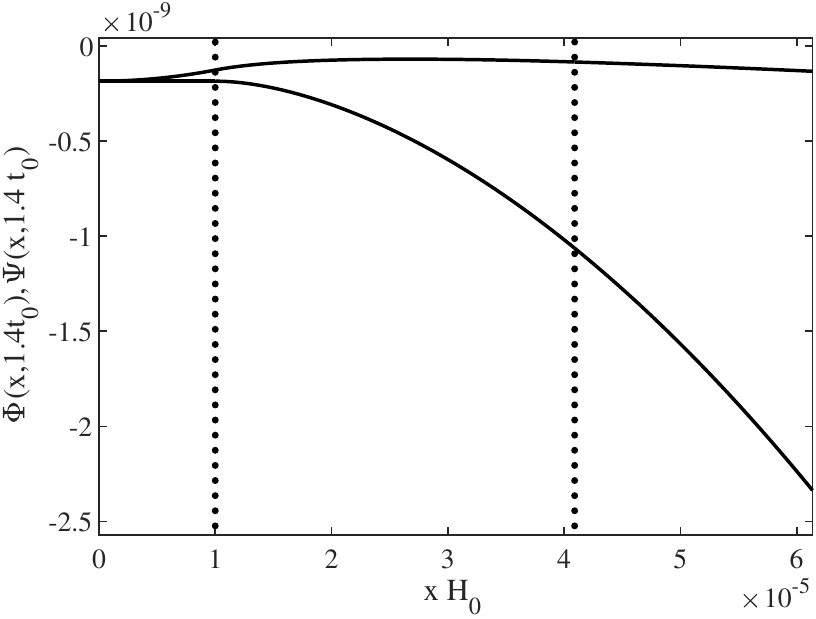}
	\caption{\changed{Shown are $\Phi$ (upper curve) and $\Psi$ (lower curve) given by equations~(\ref{eq:inphi2},\ref{eq:vacuumphi2},\ref{eq:outphi2},\ref{eq:inpsi2},\ref{eq:vacuumpsi2},\ref{eq:outpsi2}) for the same overdensity as in Figure~(\ref{fig:phi}). Coordinates are now comoving with respect to scale factor $c(t)$. Both potentials diverge quadratically on large scales since the scale factor does not correspond to the average motion there, but still remain small in a cosmological region of subhorizon size. }}
	\label{fig:phipsi}
\end{figure}

\begin{figure}
	\includegraphics[width=1.0\columnwidth]{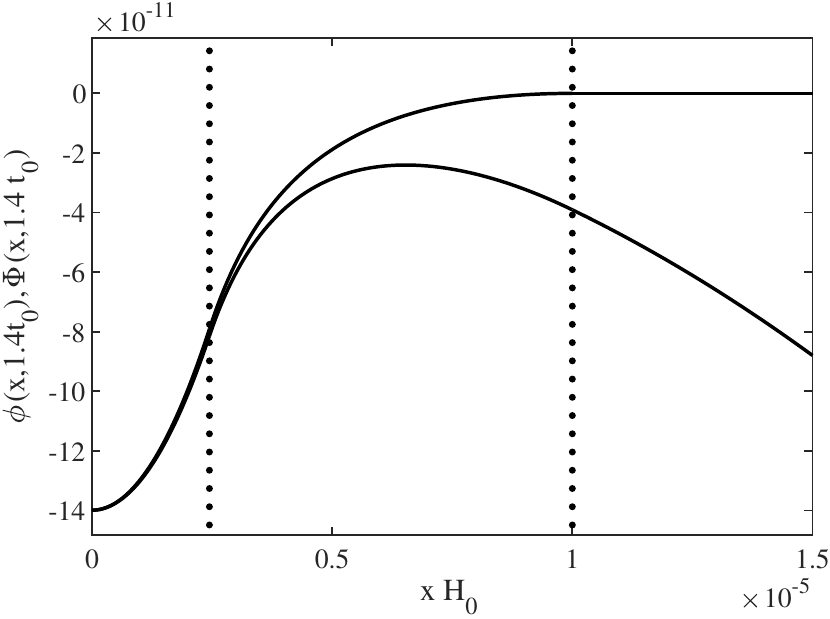}
	\caption{\changed{Comparison of $\phi$ from Figure~(\ref{fig:phi}) (upper curve) and $\Phi$ from Figure~(\ref{fig:phipsi}) (lower curve). To compare the values at the same physical points more easily, $\Phi$ has been re-expressed as a function of the comoving coordinates with respect to $b(t)$.}}
	\label{fig:phiphi}
\end{figure}

The coordinate of the top-hat boundary $x_*$ is no longer given by $b^{-1} r_*(t)$ but by $c^{-1} r_*(t) = r_0$ and is time-independent. On the other hand, the boundary of the vacuum bubble is now time-dependent with 
\begin{align} 
x_0(t) = c^{-1} r(t) = c^{-1} b \;r_0 \ge r_0. 
\end{align}
Then the solutions for the potentials are\footnote{One can recycle the previous result for $\phi$ in~(\ref{eq:inphi1}, \ref{eq:vacuumphi1}, \ref{eq:outphi1}). Compared to~(\ref{eq:poissonnormal}), at fixed time, the values of the source $\rho=T_{00}$ are the same in each region to leading order. One may rescale the terms quadratic in $x$ by $c^2/b^2$ and correct for the different Hubble rates by adding $c^2( H_b^2- H_c^2) x^2/4$. For the matching, the latter difference can be ignored because it affects all regions in the same way. Then, at the two boundaries expressed in the new coordinates, the new potential evaluates to the \emph{same} numerical value as before. The new \emph{homogeneous} solutions required for matching properly at the (new) boundaries thus need to be numerically unchanged at the boundaries as well. Therefore the \emph{constants} are unchanged (but we write $x_*(t) \rightarrow c r_0 /b$,  its previous definition, to not confuse it with the new $x_* = r_0$), and the term $\propto 1/x$ in the middle region picks up a factor of $b/c$, since $x$ at the boundary is rescaled by the same factor.}, 

	

\begin{numcases}{\Phi(x,t)=}
	 - \frac{K}{4}x^2 - \frac{3}{4} \frac{H_0^2 r_0^2}{c}   \label{eq:inphi2}   \\
	 -\frac{H_0^2 r_0^3}{2 c x} - \frac{c^2 H_c^2}{4}x^2    \label{eq:vacuumphi2}   \\
  	\frac{c^2 H_0^2 }{4 b^3}x^2 - \frac{c^2 H_c^2}{4}x^2 - \frac{3H_0^2 r_0^2}{4b}  \label{eq:outphi2} 
	\end{numcases}
\begin{numcases}{\Psi(x,t)=}
	- \frac{3}{4} \frac{H_0^2 r_0^2}{c}  \label{eq:inpsi2}  \\
	-\frac{H_0^2 r_0^3}{2c x} - \frac{H_0^2}{4 c}x^2    \label{eq:vacuumpsi2}  \\
\frac{H_0^2}{4} \left(	\frac{c^2}{ b^3}x^2 - \frac{x^2}{c}  - \frac{3 r_0^2}{b} \right)  \label{eq:outpsi2}
	\end{numcases}
where the formulas each again hold in the three regions now written as $x \leq r_0$, $r_0 < x < x_0(t)$  and $x \geq  x_0(t)$, respectively. They are shown in Figure~(\ref{fig:phipsi}). A comparison of the potentials $\Phi$ and $\phi$ from the previous section is shown in Figure~(\ref{fig:phiphi}).


\subsubsection{External region $x \ge x_0(t)$}
The discussion proceeds like in the previous case~(\ref{sec:internal1}) with $b$ and $c$ reversed in the formulas. First, the spatial coordinate is transformed analogously to~(\ref{eq:xprime}). Next, of course, $-\delta g_{tt}/2 = \Psi - A^2 b^2/2$ is a different expression, but one can again confirm that it is in fact equal to the time derivative of a $\delta t$ chosen for diagonalization of the metric as given below, so it is again possible to simultaneously let $g_{tt} \rightarrow g_{t't'} = -1$ when diagonalizing the metric. The determination of $\epsilon$ for obtaining $b^2(t')$ (instead of $b^2(t(t'))$ [without $\epsilon$] or $b^2(t(t')) (1+2\epsilon) \equiv \tilde{b}^2(t')$ [for a generic $\epsilon$]) in $g_{x'x'}$ is analogous, and all terms perturbing $b^2(t')$ cancel for this particular choice of $\epsilon$ in $g_{x' x'}$ after the transformation to the new time coordinate. Therefore, the transformation
\begin{align}
x' & =  x(1-\Phi(x,t))\left(1 - H_b(t) \delta t(x,t)\right) \frac{c(t)}{b(t)} \label{eq:trafo12} \\ 
t' &= t + \delta t(x,t) \label{eq:trafo22}  \\ 
\delta t(x,t) &\equiv \frac{H_c(t) - H_b(t)}{2} c^2(t) x^2 + g(t) \label{eq:trafo32}  \\
g(t) &\equiv -\frac{3}{4} H_0^2 r_0^2 \int b^{-1} dt \label{eq:trafo42} 
\end{align}
puts the exterior region of the perturbed FLRW metric~(\ref{eq:metricc}) with collapsing scale-factor into the EdS metric with its standard scale factor
\begin{align}
ds^2 = -dt'^2 + b^2(t')  (dx'^2  + x'^2 d\Omega^2).
\end{align}

It has to be noted that unlike before for the exterior region, this was shown only on subhorizon scales, because we have assumed smallness of the potentials which clearly both grow quadratically for this choice of scale factor. As discussed, on larger scales, the choice of a scale factor that leads to large peculiar velocities is not expected to be successful in the Newtonian limit.  

\subsubsection{Vacuum region $ r_0 \le x \le x_0(t)$ }

The transformation to the Schwarzschild line element presents no difficulties following the previous discussion in Section~(\ref{sec:vacuum1}). 
We find  
\begin{align}
r  &= x (1-\Phi(x,t)) c(t)  \\ 
t' &= t + \frac{1}{2} H_c(t) c^2(t) x^2
\end{align}
to put the perturbed FLRW metric~(\ref{eq:metricb}) in the vacuum region into the weak-field approximation of the Schwarzschild solution.

\subsubsection{Interior region $ x \le r_0$ }
The transformation is a simple special case of the general transformation from the flat FLRW metric to the curved collapsing FLRW metric~(\ref{eq:trafo1})-(\ref{eq:trafo4}) for unchanging scale factor, so for $b \rightarrow c$; non-negligible off-diagonal terms are not created and $\epsilon = H_c \delta t$  is not spatially dependent. We take
\begin{align}
x'  &= (1-\Phi(x,t))(1- H_c(t) \delta t(t)) x \\ 
t' &= t + \delta t(t)\\
\delta t(t) &= -\frac{3 H_0^2 r_0^2}{4} \int c^{-1} dt 
\end{align}
and get 
\begin{align}
g_{rr} &= c^2(t')(1 + 2 \Phi_{,x'}x') \approx c^2(t')(1+Kx'^2)^{-1} \nonumber \\ &=  c^2(t')(1-k x'^2)^{-1},
\end{align}  so that~(\ref{eq:metrictophat}) is again recovered. 

This case would have been even simpler, namely not requiring a time transformation $\delta t = 0 = \epsilon$, had we shifted the potentials by a (time-dependent) constant setting $\Psi=0$ in the inside. A similar trick was used zeroing $\phi$ in the expanding scale factor case $a(t) = b(t)$, exploited in Section~(\ref{sec:external1}). Here we have illustrated that a non-zero constant related to $\gamma(t)$ in~(\ref{eq:psiminusphi}) can be removed. 

\section{Discussion}
\label{sec:discussion}

The coordinate transformations derived in the previous section validate the perturbed FLRW description for the embedded top-hat example for different gauge choices of the scale factor for a specific example. Here we address some subtle points of the general Newtonian limit: the validity on large scales in Section~\ref{sec:emergenceofa}, the Newtonian picture of structure formation in Section~\ref{sec:expansionforce} and the relation of the \emph{linear} part of the gauge freedom $\Psi - \Phi$ in~(\ref{eq:fi}) to the Newtonian cosmology developed by Heckmann and Sch\"ucking~\citep{heckmannschuecking1955, heckmannschuecking1956} as well as implications for backreaction in the remaining two sections.

\subsection{Emergence of Friedmann scale factor}
\label{sec:emergenceofa}

On scales comparable to or larger than the Hubble scale, we have seen that the Newtonian limit can work only under a list of assumptions. 

First, we need $\Psi \simeq \Phi$ and $\alpha(t)  \simeq 0$, so we may demand  simply $\alpha(t)  = 0$ with the corresponding Friedmann solution~(\ref{eq:matterlambda}) for $a(t)$. Otherwise, if $H_a$ departs by $\mathcal{O}(H_0)$ from the Friedmann evolution, the difference of the potentials, and therefore at least one of them, approaches $\mathcal{O}(1)$ at the Hubble scale, contradicting the assumption of small potentials used to derive the Newtonian field equations in Section~\ref{sec:newtonian}. 

Second, in Section~\ref{sec:newtonian} we have also requested that the physical Hubble flow is described by $a(t)$ on large scales so that peculiar velocities can remain small far away from the coordinate center. 

These two conditions are compatible, because the Friedmann equation for $a(t)$ from the Einstein equations for $\Psi = \Phi$ and the Friedmann equation for the matter background that can be derived from the geodesic equation~(\ref{eq:NewtonianGeodesic}) (done in Section~\ref{sec:tophateqns} for the special case of vanishing cosmological constant) are in fact equivalent. 

Finally, another quadratic divergence that would spoil the Newtonian limit would arise when solving the Poisson equation~(\ref{eq:poisson}) if the source (meaning all terms except the Laplacian) did not average to something small on large scales. \changed{This is a well-known issue with the solution of the Newtonian Poisson equation~(\ref{eq:poissonnewton}), for which
\begin{align}
\phi_N \sim 2 \pi G_N/3 \, \rho_b r^2 \label{eq:quaddiv}
\end{align}
reaches the non-perturbative $\mathcal{O}(1)$ at the Hubble scale. 
The metric potential $\Phi$ differs from the Newtonian potential by a quadratic term, equation~(\ref{eq:phiphiN}), and by an appropriate choice of scale factor this regularizes all or part of the divergence. This choice is fully compatible to the previous conditions. 
}
 If we assume that at early times, when the universe was very homogeneous, the source does average to something small, it already follows from the previous assumptions that it is the case at all times: since matter moves on average with the Hubble flow described by $a(t)$, the continuity equation for matter then implies that $T_{00}$ thins out as $a^{-3}$, which is in fact proportional to the other source terms $3H_a^2-\Lambda$ in~(\ref{eq:poisson}). This can be checked explicitly from the solution~(\ref{eq:matterlambda}) for $a(t)$ obtained from $\Psi = \Phi$.

\changed{
	The smallness of the average of the source of the Poisson equation at early times becomes a statement about the flatness of the universe on large scales. If the average is not small at an early time, a curvature term may be introduced to balance the resulting divergence.
	The argument of the time-preservation of the absence of the divergence can be generalized to this case.  
}

One is of course free to introduce another scale factor and coordinate system when \emph{solving} the Newtonian equations \emph{after} they have been obtained as described, as discussed in~\citep{kaiser2017NewtonianBackreaction}. However, only the Friedmann scale factor $a(t)$ will be valid in the expression for the metric to describe the geometry of the spacetime. In this sense the $a(t)$ given by~(\ref{eq:matterlambda}) is emerging on large scales as a quantity that is useful for the description of the geometry.  


\subsection{Cosmic expansion force acting on local systems?}
\label{sec:expansionforce}

Many workers have investigated the influence of the global cosmology on the local dynamics. The classic results of~\cite{einsteinstraus1945, einsteinstraus1946} are in accordance with the more general result~(\ref{eq:NewtonianGeodesic}) reported here that there is no local effect in a matter-dominated universe that cannot be explained in a simple manner by Newtonian gravity. In particular, Newton's laws do not have to be ``corrected'' for the ``expansion of space''. 

The cosmological constant does have an effect, investigated for example in~\cite{nandra2012lambda}. More generally, any modification of the background evolution, be it homogeneous dark energy or infrared-modified gravity, replaces the cosmological constant by a term that is time dependent in general. This term can simply be understood as the Newtonian ``weight'' of the dark energy component seen as a perfect fluid, when taking as given that pressure gravitates as predicted by GR. Note that N-body simulations are sensitive to modification of the background evolution because of this mechanism.\footnote{Of course in different coordinates like comoving space and supercomoving time, employed for numerical reasons, the physics may be less obvious.} All of this certainly has been known to some cosmologists for a long time (an early qualitative discussion is~\cite{dicke1964expansion}, see also~\cite{noerdlinger1971expansion}), has been employed in studies of dynamics, e.g.~in~\cite{lahav1991lambda, falco2013dynamical}, and forms the starting point of the success of Newtonian cosmology. An insightful 
discussion of the problem can be found in~\cite{gaite2001influence}, where the Newtonian shear effects are also investigated. 

However, recent studies have sometimes come to more diffuse conclusions. These are often based on modeling the problem using exact solutions that contain a homogeneous component, like unperturbed FLRW for the the homogeneous universe~\citep{cooperstock1998influence, bonnor1999atom, price2012expanding}, exact solutions describing a mass embedded into a universe with homogeneous density component like McVittie's\footnote{Indeed, the energy-momentum tensor of McVittie's solution shows that besides the central delta function there is a homogeneous cosmic fluid everywhere.}~\citep{carrera2010influence, nandra2012mcvittie}, or others \citep{bonnor1996expansion}, or they discuss several such models~\citep{faraoni2007expansion}. 
On the other hand, the LTB solution does not contain a separate homogeneous component, and results directly agree with ours~\citep{pavlidou2014DEturnaround}.

The homogeneous background components in the universe, more precisely a ball of cosmic fluids centered in the origin and extending up to a particle's position, causes a force with magnitude proportional to $\ddot{a} r / a$ according to Newton's gravity, where $a$ is the scale factor of the background cosmology obeying the Friedmann equations. (Indeed this is a valid way to derive the Friedmann equations, see Section~\ref{sec:tophateqns} and~\citep{bertschinger1995notes, maggiore2018waves2}.) The authors investigating local systems in cosmological backgrounds based on exact solutions often report such a force, but do not seem to appreciate that it can be understood in a simple Newtonian way. This is problematic when drawing conclusions about our realistic inhomogeneous universe. One must be aware of the local distribution of sources. For example, for vanishing cosmological constant, there is no cosmic force in the solar system if and only if there is no dark matter in the solar system, and conversely a force due to dark matter can be orders of magnitude larger than $\ddot{a} r / a$ if the dark matter concentration is similarly higher. Regarding the cosmic effects on atoms, the importance of the nature of dark matter is then obvious. 

In our view, it would be a particularly unnatural interpretation to attribute the effect instead to a long-range non-Newtonian force exerted from the ``expansion of space'', which does not make sense in the Newtonian picture.  
Therefore there is no need to argue against the validity of the physical picture given by the Einstein--Straus situation, even if it is not a realistic model neglecting the hierarchy of structures and demanding an exceedingly large vacuum bubble to compensate e.g.~our sun~\citep{carrera2010influence}.

\subsection{Newtonian inertial frames in cosmology}
\label{sec:inertialframes}

In an unperturbed $\Lambda$CDM universe, the symmetry of homogeneity is incompatible with the Newtonian postulate of the existence of \emph{one particular} class of inertial frames related by Galilean transformations. Two matter particles at two different locations feel a relative force and cannot both be at rest in, and define the origin of, an inertial frame of the same class. Clearly, there is no concept of a unique, global ``absolute space'' for an infinite self-gravitating system. Even for a finite, expanding sphere there is no preferred inertial frame if observers at different locations are taken to be equivalently meaningful and ignorant of the boundary~\citep{mccrea1955Frames}. 

Newtonian cosmological theory can do without absolute space. Conceptually, one wants to adopt the local freely falling inertial frames familiar from GR. Mathematically, the breaking of the symmetry of homogeneity at the level of the specification
of the theory can be avoided and postponed to the choice of integration constants at the level of solution. This has been achieved in~\cite{heckmannschuecking1956} by completing the Newtonian Poisson equation with the following boundary conditions at infinity
\begin{align}
 \nabla^2 \phi_N &= 4 \pi G_N \rho(t)\nonumber \\ 
 \left(\partial_i \partial_j - \frac{1}{3} \delta_{ij} \nabla^2\right) \phi_N &= A_{ij} (t), \label{eq:heckmannschuecking}
\end{align}  
where $\rho(t)$ is the limit of the density at spatial infinity, assumed to exist, and $A_{ij}(t)$ is a symmetric traceless matrix function that selects between various Newtonian cosmological models. 
The solution $\phi_N$ is then unique up to the addition of planes 
\begin{align}
\phi_N \rightarrow \phi_N +  \bm{\tilde{\beta}}(t) \cdot \boldsymbol{x}+ \gamma(t).
\end{align} 
The linear term causes an additional time-dependent acceleration. This can be used to adapt the coordinates from one freely falling observer to another. More generally, any kind of \changedagain{global homogeneous acceleration can be achieved with the linear term. Unlike the effects of $A_{ij}$, however,} \changedagainagain{the linear term} is fundamentally unobservable in the present setting~\citep{heckmannschuecking1955} because all of the material content of the universe including any lab equipment is affected by it in the same way. It amounts to describing the same physics from an accelerated frame. 

In this paper, we are faced with the general-relativistic emergence of the situation captured by this modification of Newtonian theory. 
On the one hand, for the starting point of the (unperturbed) FLRW metric, the transformation from one free-fall observer to another is a trivial translation generated by the obvious spatial Killing vectors. All comoving observers are geodesic and therefore simultaneously unaccelerated from a GR point of view. On the other hand, the Newtonian limit involves the transition to expanding physical coordinates  which necessarily singles out the coordinate origin, seemingly breaking the symmetry. But the theory of GR and the FLRW ansatz on small scales with free perturbations naturally do not ``know'' that our goal is to find Newton's familiar equations written out in any inertial frame at all and certainly do not choose the origin for us.  

And indeed, GR manages to preserve the symmetry of homogeneity due to gauge freedom.\footnote{We do not specify any boundary conditions for $\Psi$. Here we assume $\Phi$ to be any fixed solution of~(\ref{eq:poisson}), the boundary conditions of which we discuss below.} The function $\bm{\beta}(t)$ is a degree of freedom arising when integrating Einstein's equations that adds a linear term $\bm{\beta}(t) \cdot \boldsymbol{x} $  to $\Psi$ (and $- 2 \bm{\beta}(t) \cdot \boldsymbol{x}$ to $g_{00}$) according to~(\ref{eq:psi}).
Again, in general, for $\bm{\beta}(t)\neq 0 $, the negative gradient generates a global acceleration $-\bm{\beta}(t)$ on a particle, as can be easily seen from the geodesic equation in the form of~(\ref{eq:accel}). 
In this case, one may boost to that proper frame which, as seen from the original one, undergoes the same acceleration. In Minkowski space, so on sufficiently small scales, such a boost precisely removes the linear term from $g_{00}$~\citep{moeller1952relativity}, replacing it by the square $(\bm{\beta}(t) \cdot \boldsymbol{x} )^2$. The effect of this remaining term in the geodesic equation must be neglected at leading order if we assume the acceleration happens on cosmic time scales $|\bm{\beta}| \sim H_0 v$. While a detailed calculation of the boost for the perturbed FLRW metric on all scales goes beyond the scope of this discussion, we may conclude that the choice of $\bm{\beta}(t)$ can be used on small scales and after taking the Newtonian limit to transform from one class of inertial frame to another, and in general it allows for the description of the same physics from an accelerated frame. Therefore, setting $\bm{\beta}(t) = 0$ (together with the assumptions in the next section) has the interpretation of selecting \emph{one} of all the mutually accelerated classes of inertial frames compatible with homogeneity as our coordinate system.

It is worth pointing out a difference between the Newtonian and GR cosmology. Unlike in the Newtonian cosmology developed in~\cite{heckmannschuecking1956}, the \changedagainagain{mathematically explicit} homogeneity of the Newtonian cosmology obtained here from GR does not rely on the engineered boundary conditions~(\ref{eq:heckmannschuecking}), but emerges naturally from the field equations.


\subsection{Boundary conditions and constraints on general-relativistic effects on the dynamics}
\label{sec:backreaction}

\changed{
In Section~\ref{sec:newtonian} we have argued that on the level of the \emph{field equations}, which are local in real space, terms nonlinear in the metric perturbations are sub-leading\footnote{Of course, throughout, we assume that a perturbative treatment is valid, where the size of the nonlinear terms is estimated using properties of the linear solution.} for late-time cosmology with relative corrections of only $\mathcal{O}(v^2)$. We have shown that the remaining Newtonian limit is fully consistent and compatible with nonlinear behaviour of matter, that is, density contrasts $\delta \gg 1$. That this picture carries over to the \emph{solutions} received support from the explicit example considered in Section~\ref{sec:tophat}. In more general settings, in the solution, due to its integral nature, small nonlinear terms in the differential equations \emph{may} accumulate and grow to become large on very large scales, if they are coherent. However, due to the suppression of the nonlinear source terms by $v^2$ with respect to the matter density, this growth is equally suppressed by a factor of $v^2$ compared to the growth of the unregularized Newtonian potential~(\ref{eq:quaddiv}), and thus bounded by 
\begin{align}
\phi_{nl} \lesssim v^2 G_N \rho_b r^2 \sim v^2 H_0^2 r^2.
\end{align} 
  On horizon scales, the scalar perturbations of the FLRW metric could thus be modified by a term of a size of up to $v^2$, comparable in magnitude to the regularized potentials. However, due to the large-scale growth, the implied force modification $|\nabla \phi_{nl}|_{r=H_0^{-1}} \sim H_0 v^2$ is still suppressed by a factor of $v$ with respect to local forces ($ |\nabla \Phi| \sim H_0 v$) and, more importantly, by a factor of $v^2$ with respect to the coherent Friedmann forces relating to the expansion (the first and the other terms in the geodesic equation~(\ref{eq:fullforce}), respectively). Thus, even in this coherent case, this is not sufficient to explain the order-unity effect of the cosmological constant on the Friedmann dynamics at horizon scale. We conclude that the nonlinear terms of the Einstein equations may really be dropped in a first-order analysis at least in a finite region of horizon size or smaller.
}

\changed{
We can further develop the discussion of possible effects from general relativity on the dynamics, under the assumption that gravitational waves can be neglected. 
Restricted to small scales, we will see that the assumption of a perturbed FLRW metric is weak enough and that we can understand the effects of cosmological solutions beyond an FLRW background despite the formal use of this metric. 
 Specifically, we ask the questions if, firstly, it is possible within the approach under consideration that ignorance of the ``right'' \emph{global} background solution could lead to an order-unity effect on the matter motion on scales large enough to have  cosmological applications, and, secondly, if this effect could look like cosmic acceleration. We will find that the answers are yes and no, respectively. 
}

\changed{
It is important to be mathematically slightly more precise in the rest of this section. To obtain a well-posed Dirichlet problem from the Poisson equation~(\ref{eq:poisson}) or its Newtonian analogue~(\ref{eq:poissonnewton}), we must prescribe boundary conditions. We can define a boundary of a finite-sized (but large) patch of the universe under consideration and prescribe the potential on it.  At first, we may ignore the gravitational effects caused by matter outside a \emph{spherical} patch and set the source to zero there; note first that in the homogeneous case the Newtonian forces from each outer spherical shell cancel exactly. In the inhomogeneous Newtonian setting the shearing caused by structures beyond $\sim 1$ Mpc, which affects only non-spherical objects, has been estimated in~\cite{gaite2001influence} to be negligibly small. 
}

\changed{
Next, as an alternative to setting boundary conditions, we can specify a Green's function that, convolved with the source, yields a unique solution for the potential, defining it also on the boundary. For Newtonian gravity, we would like this Green's function to be 
\begin{align}
G(\boldsymbol{x},\boldsymbol{x}') = - \frac{1}{4 \pi} \frac{1}{|\boldsymbol{x} - \boldsymbol{x'} |}, \label{eq:normalgreens}
\end{align}
since taking the gradient of the convolution shows that this corresponds to applying Newton's force law in the chosen patch.\footnote{This Green's function generates the isotropic Friedmann expansion in the homogeneous case. This is not happening for a non-spherical patch choice, but in this case we cannot expect the force from outside the patch to be negligible unless the total matter distribution is either finite and aspherical or the infinite limit of such a set-up, to be discussed below. 
}
}

\changed{
The effect of any different Green's function (or boundary condition) is that of adding to the solution another homogeneous solution of the Poisson equation, i.e.~a harmonic function. Among these are linear functions (planes). These have been fixed already in the previous section (e.g.~to ensure that the center of the spherical patch, where the potential gradient vanishes in the homogeneous setting, is the coordinate origin). The nonlinear harmonic functions, on the other hand, have nontrivial gradients. For example, the small effect of the matter inhomogeneities outside the spherical patch could be reintroduced by a small deformation of the Green's function. Furthermore, effects from the boundary at infinity\footnote{For example, one can consider the infinite limit of a aspherical, e.g.~elliptical, matter distribution~\cite{bertschinger1995notes}.} in Newtonian cosmology or, as we will see below, from non-FLRW background solutions in GR are encoded in boundary conditions. Similarly, if the nonlinear terms of GR were to conspire to give an effect at very large scales, this would have to appear as nontrivial boundary conditions of our patch. 
}

In the complete theory of GR  a choice of boundary conditions on a surface surrounding the region of interest is not required; providing initial data is enough.\footnote{Beyond the Newtonian approximation, which is based on elliptic constraint equations, the boundary conditions are not a free choice but fixed by the initial state and the dynamics of GR. To illustrate causality, consider GR linearized around Minkowski spacetime~\citep{maggiore2008waves1},
	$g_{\mu\nu} = \eta_{\mu\nu} + h_{\mu\nu}$ with 
	$\bar{h}_{\mu \nu} = h_{\mu\nu} - \eta_{\mu\nu} h /2$ : in Lorenz gauge (which is exhibiting causality more explicitly) 
	$\partial_\mu \bar{h}_{\mu \nu} = 0$ 
	one has 
	$\Box \bar{h}_{\mu \nu} = - 16 \pi G_N T_{\mu \nu}$ 
	and therefore 
	\begin{align}
	-\frac{1}{4} \bar{h}_{00} = -G_N \int d^3x' \frac{T_{00}(t - |\boldsymbol{x} - \boldsymbol{x'} |, \boldsymbol{x'})}{|\boldsymbol{x} - \boldsymbol{x'} |} \label{eq:linsol}
	\end{align}
	where the retarded Green's function has been selected to causally propagate the initial state to the future. 
	When the retardation and the other components of the source are neglected, (\ref{eq:linsol}) becomes the Newtonian potential obtained using the Green's function~(\ref{eq:normalgreens})
	\begin{align}
	-\frac{1}{4} \bar{h}_{00} = -\frac{1}{2} h_{00} = -\frac{1}{2} h_{i i}  = -G_N \int d^3x' \frac{T_{00}(t, \boldsymbol{x'})}{|\boldsymbol{x} - \boldsymbol{x'} |} \\
	= 4 \pi G_N \int d^3x' G(\boldsymbol{x}, \boldsymbol{x'}) T_{00}(t, \boldsymbol{x'}) 
	\end{align} of the Laplacian, which fixes the value of the potential on any boundary. Similarly, in a standard global approach to cosmology, the boundary conditions in the Newtonian approximation are at least in principle fixed from GR assuming (almost) homogeneous FLRW at early times.}
\changed{ However, after taking the Newtonian limit, we have lost the information of the evolution of the boundaries. Thus, while the Green's function is actually determined and cannot just be chosen to be~(\ref{eq:normalgreens}), we are ignorant of it, also because we do not want to control the initial state in our approach on very large scales, preferring to restrict linearization to a subhorizon patch. 
Therefore, we simply consider the possibility of generic boundary conditions which can even be time-dependent. The only constraint we put on them is that they cause effects that are at most of order unity, that is, they cause additional forces comparable to the Newtonian forces, since even stronger effects are certainly not observed. Equivalently, we are allowed to add additional harmonic functions $\phi_h$ to the potential that have gradients comparable to those of the Newtonian potential obtained with~(\ref{eq:normalgreens}).  }

\changed{ It follows that these functions $\phi_h$ do not grow arbitrarily fast. Since we want $| \nabla \phi_h | \lesssim  H_0 v + H_0^2 r$ (again corresponding to the local forces and the Friedmann forces, respectively, equation~(\ref{eq:fullforce})), they reach at most $\mathcal{O}(1) $ at the horizon scale. Upon subtraction of an irrelevant constant, we therefore have $\phi_h \sim v^2 $ on sub-horizon scales $\lesssim v/H_0$. This scale is already cosmological in the sense that the average matter density in a region of this size can be comparable to that of the cosmological constant, and the latter therefore effective in causing an accelerated Friedmann evolution of the patch. The approximations of Section~\ref{sec:newtonian} still hold. Furthermore, the choice of scale factor is irrelevant on this scale as discussed in Section~\ref{sec:friedmannasgauge} (c.f.~Section~\ref{sec:emergenceofa}); in particular $a=1$ is allowed, corresponding to an expansion around Minkowski spacetime\footnote{Note that for this case, and for $\Lambda = 0$, the usual $\Psi = \Phi$ follows from equation~(\ref{eq:psiminusphi}), with $a=1$ corresponding to the solution of the second-order Friedmann equation~(\ref{eq:Friedmann3}) for $H_a(t_0) = 0$.}. It is clear that there is no additional input from the fact that the starting point was a perturbed FLRW metric. The situation is instead that locally a given metric, in fact general up to neglecting tensor modes and smallness of vector modes, satisfies Einstein's equations for a given source to some required accuracy under the assumptions of Section~\ref{sec:newtonian}, and uniqueness of the solution for given initial conditions implies that we are considering the right solution if the initial conditions are satisfied. Since we have encoded the nontrivial initial conditions in the scalar sector into the boundary conditions, which are still free, we are facing the general set of all relevant scalar sector solutions of GR for the local cosmological patch with a given dust source. In particular, as mentioned, there is no assumptions on isotropy at this point (within the limits of order-unity effects). As another example, locally, curvature of spatial slices can be realized perturbatively, as is clear from the example in Section~\ref{sec:tophat}, where the coordinate transformation between the curved FLRW metric for the collapsing core and the flat FLRW metric description was given. We conclude that on cosmological but subhorizon scales, the perturbed FLRW metric is a fully general ansatz for the relevant physics that does not imply a choice of background. On the other hand, we cannot trivially extend such a local treatment to horizon scales, because $\phi_h$ can become nonperturbative, just like $\Phi$ and $\Psi-\Phi$ for the wrong choice of metric scale factor. We have thus made the assumption of a FLRW background earlier when claiming that can $\Psi$ and $\Phi$ remain small on large scales, making an implicit statement on their boundary conditions. 
}


Despite the generality of the free boundary conditions, they cannot cause radial forces like the cosmic forces, including the accelerating one caused by a cosmological constant. To see this, note that the gradients of the harmonic functions $\phi_h$ are necessarily divergence-free. Therefore, by Gauss' theorem, their flux through any closed surface must vanish, which is impossible for a radial force field. \changedagainagain{The gradients instead correspond to anisotropic (multipole) forces causing tidal shear. As mentioned above, the shear due to the inhomogeneity of the matter outside the patch is normally negligible in our universe~\citep{gaite2001influence}. However, 
one can conceive cosmological models that are not based on an isotropic and homogeneous FLRW background. These effects can be modeled locally by nonlinear harmonic functions $\phi_h$ since we have seen that this is the only remaining degree of freedom. It is interesting to note that these functions are in one-to-one correspondence with the different solutions of Heckmann-Sch\"ucking Newtonian cosmology selected by different boundary conditions~(\ref{eq:heckmannschuecking}) because those boundary conditions are precisely such that the solution to the Poisson equation is unique up to the addition of planes~\citep{heckmannschuecking1956} and therefore they precisely fix the nonlinear functions $\phi_h$. Consequently, these general anisotropic Newtonian models, as well as their exact relativistic counterparts with nontrivial electric part of the Weyl tensor~\citep{ellis2009relcosmo} in the Newtonian limit, admit a perturbed FLRW description by using some Green's function different from~(\ref{eq:normalgreens}) on subhorizon scales. }

\changed{
Conversely, if anisotropies are sufficiently constrained (which seems natural given the observed isotropy of the cosmic microwave background), 
there cannot be large boundary conditions that break our perturbative treatment. The relevant solutions of GR are then restricted to be close to the FLRW type even on larger scales. We reserve a quantitative version of this argument for future work. }

For a much more rigorous and more general discussion of backreaction, see~\cite{ishibashiwald2005, greenwald2011,  green2013examples, green2014well}.

\section{Summary and Conclusion}
\label{sec:conclusion}

We have \changed{justified and} analyzed the Newtonian limit of late-time GR cosmology with a perturbed FLRW ansatz including scalar and vector modes, \changed{following an approach in real space and implementing assumptions on the matter structures and motion based on observations.} The vector mode of the metric can be used to absorb the transverse component of the time-space components of the energy-momentum tensor and does not enter elsewhere to leading order. The dynamics of cold dark matter is Newtonian to leading order, \changed{which was discussed mostly assuming flatness and isotropy, although generalizations have been pointed out}. The scale factor in the metric, or equivalently the spatial curvature at some instance, is a gauge choice on subhorizon scales and always drops out of the dynamical equations when they are formulated in a Newtonian frame. GR does not select the inertial frame; selecting it is another gauge choice. This can be understood as a consequence of the cosmological principle encoded in the FLRW geometry, and its implementation into Newtonian cosmology has previously been achieved in the Heckmann-Sch\"ucking formulation of the boundary conditions. 

On large scales comparable or beyond the Hubble scale, given natural assumptions, the Newtonian limit works, but the scale factor appearing in the metric must be chosen to be close enough to the standard one compatible with the Friedmann equations with the right spatial curvature. \changed{In Fourier space, on horizon scales, one finds general-relativistic corrections which however correspond to Newtonian cosmology in changed variables~\citep{greenwald2012newtonian}. Since we work in real space to define accuracy, we are not concerned with these terms in this work}.

\changed{Restricting to small vector perturbations and ignoring gravitational waves, GR does not allow for other dust cosmologies than those of Heckmann and Sch\"ucking~\citep{heckmannschuecking1956} on subhorizon scales. Therefore, comparing observations of structure on these scales to predictions derived with the perturbed FLRW metric are actually general if, in case it is of interest, anisotropic boundary conditions are allowed.}

\changed{
	Consequently, measuring the cosmological dynamics on subhorizon scales can directly probe and disentangle the existence and nature of dark energy versus relativistic dynamical effects (encoded in boundary conditions). For example, one may study properties of galaxy clusters like the turn-around radius, see~\cite{farbodturnaorund} and references therein. Less directly, one may restrict comparing clustering data to predictions based on the perturbed FLRW approach with various kinds of dark energy to sub-horizon scales to restore model-independence. On the other hand, supernova measurements of the expansion history have to extend to large distances of redshifts $z\sim 1$ to make the effects of dark energy on the cosmological background dynamics visible, by going sufficiently far into the past on the light cone instead of on particle trajectories, exceeding the sub-horizon regime. On these larger scales, a more advanced treatment including GR cosmologies beyond FLRW backgrounds would be required to make the most of the data in particular if GR supports cosmologies that are both relevant for our universe and not simple large-scale extensions of the Heckmann-Sch\"ucking Newtonian cosmologies, such that new radial forces can emerge on large scales. However, this seems difficult to achieve without simultaneously violating observed isotropy already on smaller scales. }

Furthermore, we fully agree with the Newtonian picture in~\cite{kaiser2017NewtonianBackreaction} on the distinction of volume expansion and a scale factor that is a gauge-freedom, and share the critique of the N-body simulation carried out in~\cite{racz2017nbody}, where the two concepts were mixed\changed{, and the metric scale factor was locally adjusted to the expansion rate. We have shown that changing the metric scale factor in a patch must be accompanied by a change of metric curvature according to Einstein's equations, and this in fact ensures that the Newtonian limit, which holds to leading order, is unchanged. Failing to do so simulates not relativistic physics beyond a FLRW background, but another theory than GR.}

We conclude that the perturbed FLRW ansatz provides a consistent physical picture. It works as well as one could possibly hope for for the embedded top-hat model we have studied, for which the exact solution is known. From this analysis there are no concerns for using the perturbed FLRW metric to simultaneously describe cosmology and much smaller scales like the solar system. 

The equations of motion for matter can be generalized e.g.~to include frame dragging from the vector mode. In~\cite{castiblanco2018RelLSS} the authors claim that the relativistic dynamical corrections beyond the Newtonian limit can have a very significant effect on the squeezed bispectrum due to the combination of different scales, at least when evaluated in perturbation theory. If confirmed by nonperturbative numerical studies, this would limit the success of Newtonian cosmology (amended with relativistic light propagation) in the perturbed FLRW metric~(\ref{eq:lineelement}) for late-time LSS.

\section*{Acknowledgements} 
I thank Michele Maggiore, Pierre Fleury and Enis Belgacem for many crucial discussions and encouraging support and Pierre Fleury for a careful reading of the manuscript and insightful comments. Ruth Durrer, Steen Hansen, Nick Kaiser, David F. Mota and Cl\'ement Stahl,  \changedagain{as well as the anonymous referee,} have also provided well-appreciated feedback on various aspects of this work.  
\changed{ My work is supported by the Fonds National Suisse and by the SwissMAP National Center
	of Competence in Research.}


\bibliographystyle{mnras}
\bibliography{references}

\begin{thebibliography}{}
\makeatletter
\relax
\def\mn@urlcharsother{\let\do\@makeother \do\$\do\&\do\#\do\^\do\_\do\%\do\~}
\def\mn@doi{\begingroup\mn@urlcharsother \@ifnextchar [ {\mn@doi@}
  {\mn@doi@[]}}
\def\mn@doi@[#1]#2{\def\@tempa{#1}\ifx\@tempa\@empty \href
  {http://dx.doi.org/#2} {doi:#2}\else \href {http://dx.doi.org/#2} {#1}\fi
  \endgroup}
\def\mn@eprint#1#2{\mn@eprint@#1:#2::\@nil}
\def\mn@eprint@arXiv#1{\href {http://arxiv.org/abs/#1} {{\tt arXiv:#1}}}
\def\mn@eprint@dblp#1{\href {http://dblp.uni-trier.de/rec/bibtex/#1.xml}
  {dblp:#1}}
\def\mn@eprint@#1:#2:#3:#4\@nil{\def\@tempa {#1}\def\@tempb {#2}\def\@tempc
  {#3}\ifx \@tempc \@empty \let \@tempc \@tempb \let \@tempb \@tempa \fi \ifx
  \@tempb \@empty \def\@tempb {arXiv}\fi \@ifundefined
  {mn@eprint@\@tempb}{\@tempb:\@tempc}{\expandafter \expandafter \csname
  mn@eprint@\@tempb\endcsname \expandafter{\@tempc}}}

\bibitem[\protect\citeauthoryear{Adamek, Daverio, Durrer  \& Kunz}{Adamek
  et~al.}{2013}]{adamek2013perturb}
Adamek J.,  Daverio D.,  Durrer R.,   Kunz M.,  2013, Physical Review D, 88,
  103527

\bibitem[\protect\citeauthoryear{Adamek, Daverio, Durrer  \& Kunz}{Adamek
  et~al.}{2016}]{adamek2016gevolution}
Adamek J.,  Daverio D.,  Durrer R.,   Kunz M.,  2016, Nature physics, 12, 346

\bibitem[\protect\citeauthoryear{Aghanim et~al.,}{Aghanim
  et~al.}{2018}]{planck2018}
Aghanim N.,  et~al., 2018, arXiv preprint arXiv:1807.06209

\bibitem[\protect\citeauthoryear{Balbinot, Bergamini  \& Comastri}{Balbinot
  et~al.}{1988}]{balbinot1988einsteinstraus}
Balbinot R.,  Bergamini R.,   Comastri A.,  1988, Physical Review D, 38, 2415

\bibitem[\protect\citeauthoryear{Baumann, Nicolis, Senatore  \&
  Zaldarriaga}{Baumann et~al.}{2012}]{baumann2012EFTLSS}
Baumann D.,  Nicolis A.,  Senatore L.,   Zaldarriaga M.,  2012, Journal of
  Cosmology and Astroparticle Physics, 2012, 051

\bibitem[\protect\citeauthoryear{Belgacem, Finke, Frassino  \&
  Maggiore}{Belgacem et~al.}{2019}]{me2019LLR}
Belgacem E.,  Finke A.,  Frassino A.,   Maggiore M.,  2019, Journal of
  Cosmology and Astroparticle Physics, 2019, 035

\bibitem[\protect\citeauthoryear{Bertschinger}{Bertschinger}{1995}]{bertschinger1995notes}
Bertschinger E.,  1995, arXiv preprint astro-ph/9503125

\bibitem[\protect\citeauthoryear{Bonnor}{Bonnor}{1996}]{bonnor1996expansion}
Bonnor W.,  1996, Monthly Notices of the Royal Astronomical Society, 282, 1467

\bibitem[\protect\citeauthoryear{Bonnor}{Bonnor}{1999}]{bonnor1999atom}
Bonnor W.~B.,  1999, Classical and Quantum Gravity, 16, 1313

\bibitem[\protect\citeauthoryear{Buchert}{Buchert}{2017}]{buchert2017backreaction}
Buchert T.,  2017, Monthly Notices of the Royal Astronomical Society: Letters,
  473, L46

\bibitem[\protect\citeauthoryear{Buchert \& R{\"a}s{\"a}nen}{Buchert \&
  R{\"a}s{\"a}nen}{2012}]{buchert2012backreaction}
Buchert T.,  R{\"a}s{\"a}nen S.,  2012, Annual Review of Nuclear and Particle
  Science, 62, 57

\bibitem[\protect\citeauthoryear{Carrera \& Giulini}{Carrera \&
  Giulini}{2010}]{carrera2010influence}
Carrera M.,  Giulini D.,  2010, Reviews of Modern Physics, 82, 169

\bibitem[\protect\citeauthoryear{Castiblanco, Gannouji, Nore{\~n}a  \&
  Stahl}{Castiblanco et~al.}{2019}]{castiblanco2018RelLSS}
Castiblanco L.,  Gannouji R.,  Nore{\~n}a J.,   Stahl C.,  2019, Journal of
  Cosmology and Astroparticle Physics, 2019, 030

\bibitem[\protect\citeauthoryear{Cooperstock, Faraoni  \& Vollick}{Cooperstock
  et~al.}{1998}]{cooperstock1998influence}
Cooperstock F.~I.,  Faraoni V.,   Vollick D.~N.,  1998, The Astrophysical
  Journal, 503, 61

\bibitem[\protect\citeauthoryear{Dai, Pajer  \& Schmidt}{Dai
  et~al.}{2015}]{dai2015separateuniverses}
Dai L.,  Pajer E.,   Schmidt F.,  2015, Journal of Cosmology and Astroparticle
  Physics, 2015, 059

\bibitem[\protect\citeauthoryear{Dicke \& Peebles}{Dicke \&
  Peebles}{1964}]{dicke1964expansion}
Dicke R.~H.,  Peebles P. J.~E.,  1964, Physical Review Letters, 12, 435

\bibitem[\protect\citeauthoryear{Dominguez \& Gaite}{Dominguez \&
  Gaite}{2001}]{gaite2001influence}
Dominguez A.,  Gaite J.,  2001, EPL (Europhysics Letters), 55, 458

\bibitem[\protect\citeauthoryear{Durrer}{Durrer}{2008}]{durrer2008CMB}
Durrer R.,  2008, The Cosmic Microwave Background, by Ruth Durrer. ISBN
  978-0-521-84704-9 (HB). Published by Cambridge University Press, Cambridge,
  UK, 2008.

\bibitem[\protect\citeauthoryear{Ehlers \& Sachs}{Ehlers \&
  Sachs}{1969}]{ehlerssachs1968}
Ehlers J.,  Sachs R.,  1969, in Chretien M.,  Deser S.,   Goldstein J.,  eds,
  Vol. 2, Brandeis University Summer Institute In Theoretical Physics 1968,
  Astrophysics and General Relativity. Gordon and Breach Science Publishers,
  New York, pp 331--383

\bibitem[\protect\citeauthoryear{Einstein \& Straus}{Einstein \&
  Straus}{1945}]{einsteinstraus1945}
Einstein A.,  Straus E.~G.,  1945, Reviews of Modern Physics, 17, 120

\bibitem[\protect\citeauthoryear{Einstein \& Straus}{Einstein \&
  Straus}{1946}]{einsteinstraus1946}
Einstein A.,  Straus E.,  1946, Reviews of Modern Physics, 18, 148

\bibitem[\protect\citeauthoryear{Ellis}{Ellis}{2009}]{ellis2009relcosmo}
Ellis G.~F.,  2009, General Relativity and Gravitation, 41, 581

\bibitem[\protect\citeauthoryear{Falco, Mamon, Wojtak, Hansen  \&
  Gottl{\"o}ber}{Falco et~al.}{2013}]{falco2013dynamical}
Falco M.,  Mamon G.~A.,  Wojtak R.,  Hansen S.~H.,   Gottl{\"o}ber S.,  2013,
  Monthly Notices of the Royal Astronomical Society, 436, 2639

\bibitem[\protect\citeauthoryear{Faraoni \& Jacques}{Faraoni \&
  Jacques}{2007}]{faraoni2007expansion}
Faraoni V.,  Jacques A.,  2007, Physical Review D, 76, 063510

\bibitem[\protect\citeauthoryear{Fidler, Tram, Rampf, Crittenden, Koyama  \&
  Wands}{Fidler et~al.}{2017}]{fidler2017weakfieldNbody}
Fidler C.,  Tram T.,  Rampf C.,  Crittenden R.,  Koyama K.,   Wands D.,  2017,
  Journal of Cosmology and Astroparticle Physics, 2017, 022

\bibitem[\protect\citeauthoryear{Fienga, Laskar, Exertier, Manche  \&
  Gastineau}{Fienga et~al.}{2015}]{fienga2015}
Fienga A.,  Laskar J.,  Exertier P.,  Manche H.,   Gastineau M.,  2015,
  Celestial Mechanics and Dynamical Astronomy, 123, 325

\bibitem[\protect\citeauthoryear{Goldberg, Clifton  \& Malik}{Goldberg
  et~al.}{2017}]{goldberg2017allscales}
Goldberg S.~R.,  Clifton T.,   Malik K.~A.,  2017, Physical Review D, 95,
  043503

\bibitem[\protect\citeauthoryear{Green \& Wald}{Green \&
  Wald}{2011}]{greenwald2011}
Green S.~R.,  Wald R.~M.,  2011, Physical Review D, 83, 084020

\bibitem[\protect\citeauthoryear{Green \& Wald}{Green \&
  Wald}{2012}]{greenwald2012newtonian}
Green S.~R.,  Wald R.~M.,  2012, Physical Review D, 85, 063512

\bibitem[\protect\citeauthoryear{Green \& Wald}{Green \&
  Wald}{2013}]{green2013examples}
Green S.~R.,  Wald R.~M.,  2013, Physical Review D, 87, 124037

\bibitem[\protect\citeauthoryear{Green \& Wald}{Green \&
  Wald}{2014}]{green2014well}
Green S.~R.,  Wald R.~M.,  2014, Classical and Quantum Gravity, 31, 234003

\bibitem[\protect\citeauthoryear{Hansen, Hassani, Lombriser  \& Kunz}{Hansen
  et~al.}{2019}]{farbodturnaorund}
Hansen S.~H.,  Hassani F.,  Lombriser L.,   Kunz M.,  2019

\bibitem[\protect\citeauthoryear{Harrison}{Harrison}{1967}]{harrison1967normal}
Harrison E.,  1967, Reviews of Modern Physics, 39, 862

\bibitem[\protect\citeauthoryear{Heckmann \& Sch{\"u}cking}{Heckmann \&
  Sch{\"u}cking}{1955}]{heckmannschuecking1955}
Heckmann O.,  Sch{\"u}cking E.,  1955, Zeitschrift fur Astrophysik, 38, 95

\bibitem[\protect\citeauthoryear{Heckmann \& Sch{\"u}cking}{Heckmann \&
  Sch{\"u}cking}{1956}]{heckmannschuecking1956}
Heckmann O.,  Sch{\"u}cking E.,  1956, Zeitschrift fur Astrophysik, 40, 81

\bibitem[\protect\citeauthoryear{Hofmann \& M{\"u}ller}{Hofmann \&
  M{\"u}ller}{2018}]{hofmann2018LLR}
Hofmann F.,  M{\"u}ller J.,  2018, Classical and Quantum Gravity, 35, 035015

\bibitem[\protect\citeauthoryear{Holz \& Wald}{Holz \&
  Wald}{1998}]{holz1998Lensing}
Holz D.~E.,  Wald R.~M.,  1998, Physical Review D, 58, 063501

\bibitem[\protect\citeauthoryear{Ishibashi \& Wald}{Ishibashi \&
  Wald}{2005}]{ishibashiwald2005}
Ishibashi A.,  Wald R.~M.,  2005, Classical and Quantum Gravity, 23, 235

\bibitem[\protect\citeauthoryear{Jelic-Cizmek, Lepori, Adamek  \&
  Durrer}{Jelic-Cizmek et~al.}{2018}]{jelic2018vorticity}
Jelic-Cizmek G.,  Lepori F.,  Adamek J.,   Durrer R.,  2018, Journal of
  Cosmology and Astroparticle Physics, 2018, 006

\bibitem[\protect\citeauthoryear{Kaiser}{Kaiser}{2017}]{kaiser2017NewtonianBackreaction}
Kaiser N.,  2017, Monthly Notices of the Royal Astronomical Society, 469, 744

\bibitem[\protect\citeauthoryear{Kim, Lasenby  \& Hobson}{Kim
  et~al.}{2018}]{kim2018spherically}
Kim D.~Y.,  Lasenby A.~N.,   Hobson M.~P.,  2018, General Relativity and
  Gravitation, 50

\bibitem[\protect\citeauthoryear{Kolb, Marra  \& Matarrese}{Kolb
  et~al.}{2010}]{kolb2010backreaction}
Kolb E.~W.,  Marra V.,   Matarrese S.,  2010, General Relativity and
  Gravitation, 42, 1399

\bibitem[\protect\citeauthoryear{Laarakkers \& Poisson}{Laarakkers \&
  Poisson}{2001}]{laarakkers2001einsteinstraus}
Laarakkers W.~G.,  Poisson E.,  2001, Physical Review D, 64, 084008

\bibitem[\protect\citeauthoryear{Lahav, Lilje, Primack  \& Rees}{Lahav
  et~al.}{1991}]{lahav1991lambda}
Lahav O.,  Lilje P.~B.,  Primack J.~R.,   Rees M.~J.,  1991, Monthly Notices of
  the Royal Astronomical Society, 251, 128

\bibitem[\protect\citeauthoryear{Maggiore}{Maggiore}{2008}]{maggiore2008waves1}
Maggiore M.,  2008, Gravitational Waves: Volume 1: Theory and Experiments.
 Vol. 1, Oxford university press

\bibitem[\protect\citeauthoryear{Maggiore}{Maggiore}{2018}]{maggiore2018waves2}
Maggiore M.,  2018, Gravitational Waves: Volume 2: Astrophysics and Cosmology.
 Vol. 2, Oxford University Press

\bibitem[\protect\citeauthoryear{McCrea}{McCrea}{1955}]{mccrea1955Frames}
McCrea W.,  1955, The Mathematical Gazette, 39, 287

\bibitem[\protect\citeauthoryear{McVittie}{McVittie}{1933}]{mcvittie1933}
McVittie G.~C.,  1933, Monthly Notices of the Royal Astronomical Society, 93,
  325

\bibitem[\protect\citeauthoryear{Milillo, Bertacca, Bruni  \& Maselli}{Milillo
  et~al.}{2015}]{milillo2015link}
Milillo I.,  Bertacca D.,  Bruni M.,   Maselli A.,  2015, Physical Review D,
  92, 023519

\bibitem[\protect\citeauthoryear{Misner, Thorne, Wheeler  \& Kaiser}{Misner
  et~al.}{2017}]{MTW2017collapsestar}
Misner C.~W.,  Thorne K.~S.,  Wheeler J.~A.,   Kaiser D.~I.,  2017,
  Gravitation.
Princeton University Press

\bibitem[\protect\citeauthoryear{M\o{}ller}{M\o{}ller}{1952}]{moeller1952relativity}
M\o{}ller C.,  1952, The Theory of Relativity.
Oxford University Press

\bibitem[\protect\citeauthoryear{Nandra, Lasenby  \& Hobson}{Nandra
  et~al.}{2012a}]{nandra2012mcvittie}
Nandra R.,  Lasenby A.~N.,   Hobson M.~P.,  2012a, Monthly Notices of the Royal
  Astronomical Society, 422, 2931

\bibitem[\protect\citeauthoryear{Nandra, Lasenby  \& Hobson}{Nandra
  et~al.}{2012b}]{nandra2012lambda}
Nandra R.,  Lasenby A.~N.,   Hobson M.~P.,  2012b, Monthly Notices of the Royal
  Astronomical Society, 422, 2945

\bibitem[\protect\citeauthoryear{Noerdlinger \& Petrosian}{Noerdlinger \&
  Petrosian}{1971}]{noerdlinger1971expansion}
Noerdlinger P.~D.,  Petrosian V.,  1971, The Astrophysical Journal, 168, 1

\bibitem[\protect\citeauthoryear{Nottale}{Nottale}{1982}]{nottale1982tophat}
Nottale L.,  1982, Astronomy and Astrophysics, 110, 9

\bibitem[\protect\citeauthoryear{Oppenheimer \& Snyder}{Oppenheimer \&
  Snyder}{1939}]{oppenheimersnyder1939}
Oppenheimer J.~R.,  Snyder H.,  1939, Physical Review, 56, 455

\bibitem[\protect\citeauthoryear{Pavlidou, Tetradis  \& Tomaras}{Pavlidou
  et~al.}{2014}]{pavlidou2014DEturnaround}
Pavlidou V.,  Tetradis N.,   Tomaras T.,  2014, Journal of Cosmology and
  Astroparticle Physics, 2014, 017

\bibitem[\protect\citeauthoryear{Peebles}{Peebles}{1980}]{peebles1980large}
Peebles P. J.~E.,  1980, The large-scale structure of the universe.
Princeton university press

\bibitem[\protect\citeauthoryear{Pitjeva \& Pitjev}{Pitjeva \&
  Pitjev}{2013}]{pitjeva2013}
Pitjeva E.,  Pitjev N.,  2013, Monthly Notices of the Royal Astronomical
  Society, 432, 3431

\bibitem[\protect\citeauthoryear{Price \& Romano}{Price \&
  Romano}{2012}]{price2012expanding}
Price R.~H.,  Romano J.~D.,  2012, American Journal of Physics, 80, 376

\bibitem[\protect\citeauthoryear{R{\'a}cz, Dobos, Beck, Szapudi  \&
  Csabai}{R{\'a}cz et~al.}{2017}]{racz2017nbody}
R{\'a}cz G.,  Dobos L.,  Beck R.,  Szapudi I.,   Csabai I.,  2017, Monthly
  Notices of the Royal Astronomical Society: Letters, 469, L1

\bibitem[\protect\citeauthoryear{R{\"a}s{\"a}nen}{R{\"a}s{\"a}nen}{2010}]{rasanen2010FLRW}
R{\"a}s{\"a}nen S.,  2010, Physical Review D, 81, 103512

\bibitem[\protect\citeauthoryear{Sanghai \& Clifton}{Sanghai \&
  Clifton}{2015}]{sanghai2015}
Sanghai V. A.~A.,  Clifton T.,  2015, \mn@doi [Phys. Rev. D]
  {10.1103/PhysRevD.91.103532}, 91, 103532

\bibitem[\protect\citeauthoryear{Sch{\"u}cking}{Sch{\"u}cking}{1954}]{schuecking1954}
Sch{\"u}cking E.,  1954, Zeitschrift f{\"u}r Physik, 137, 595

\bibitem[\protect\citeauthoryear{Thomas, Bruni  \& Wands}{Thomas
  et~al.}{2015}]{thomas2015vectors}
Thomas D.~B.,  Bruni M.,   Wands D.,  2015, Monthly Notices of the Royal
  Astronomical Society, 452, 1727

\bibitem[\protect\citeauthoryear{Van~Acoleyen}{Van~Acoleyen}{2008}]{karel2008LTBFLRW}
Van~Acoleyen K.,  2008, Journal of Cosmology and Astroparticle Physics, 2008,
  028

\bibitem[\protect\citeauthoryear{Wiltshire}{Wiltshire}{2011}]{wiltshire2011dust}
Wiltshire D.~L.,  2011, Classical and Quantum Gravity, 28, 164006

\bibitem[\protect\citeauthoryear{Yamamoto, Marra, Mukhanov  \& Sasaki}{Yamamoto
  et~al.}{2016}]{yamamoto2016LTBFLRW}
Yamamoto K.,  Marra V.,  Mukhanov V.,   Sasaki M.,  2016, Journal of Cosmology
  and Astroparticle Physics, 2016, 030

\makeatother
\end{thebibliography}

\appendix

\section{Normal coordinates for Einstein--de Sitter spacetime}

We can apply the methods developed above to the \emph{un}perturbed FLRW metric. After going to $r = a(t) x$ we have  $A= Hr, \; B = 1$ and 
\begin{align}
ds^2 = -(1-H^2 r^2 ) dt^2 - 2 H r dr dt +  dr^2 + r^2 d\Omega^2, 
\end{align}
which is exact. The most general exact diagonalizing transformation for the Einstein--de Sitter case is now $t' = f(t,r) = F(9 t^{2/3} /2 + r^2 t^{-4/3})$; to get units of time we take $F(x) = (2 x/9)^{3/2}$ such that $t' = t (1+H^2 r^2/2)^{3/2}$. This of course again reduces to $t' = t + H r^2/2$ in the subhorizon limit. In any case, $g_{rr} \rightarrow 1 + H^2 r^2/(1-H^2 r^2) = 1/(1-H^2 r^2)$ . A short calculation gives $g_{tt} \rightarrow -((1+H^2r^2/2)(1-H^2 r^2))^{-1}$. Therefore
\begin{align}
ds^2 = -&\frac{1}{(1+H^2r^2/2)(1-H^2 r^2)} dt'^2\\ &\qquad + \frac{1}{1-H^2 r^2} dr^2 \;+ r^2 d\Omega^2. \nonumber
\end{align}
Note that the time $t$  in $H \propto t^{-1}$ has to be expressed via $t'$ and $r$ from by solving the relation $t' = f(t,r)$ for $t$, and in terms of the new coordinates the metric appears very complicated. However, at the coordinate origin, the Christoffel symbols agree with those of flat spacetime in spherical coordinates, since there are no linear corrections in $r$, and so the coordinates agree with Riemann normal coordinates for the central observer up to second order.

For a general FLRW metric (not necessarily matter-dominated) and this time approximating we still find $f(r,t) = t + H r^2/2 + h(t)$, thus $g_{rr} \approx  1 + H^2 r^2$. The time-time component becomes $g_{t't'} \approx -(1-H^2r^2)/(1+ \dot{H} r^2/2 + \dot{h})^2$. For $h=0$, this is the weak-field de Sitter metric for $\dot{H} = 0$; if on the other hand $\dot{H} = -3 H^2 /2$ for Einstein--de Sitter,  $g_{tt} \approx (1+H^2r^2/2)$ agrees with the weak-field limit of the previous exact result.

\bsp	
\label{lastpage}
\end{document}